\documentclass[preprint,showpacs,preprintnumbers,amsmath,amssymb,
nofootinbib,eqsecnum,12pt]{revtex4}

\usepackage{graphicx}

\newlength{\dummysp}
\newcommand{\beq}{\begin{eqnarray}}
\newcommand{\eeq}{\end{eqnarray}}

\newcommand{\p}{{\partial}}
\newcommand{\e}{{\epsilon}}

\newcommand{\vev}[1]{{\langle #1 \rangle}}

\newcommand{\ord}[1]{{{\cal O}(#1)}}
\newcommand{\gappeq}{\mathrel{\rlap {\raise.5ex\hbox{$>$}}
{\lower.5ex\hbox{$\sim$}}}}
\newcommand{\lappeq}{\mathrel{\rlap{\raise.5ex\hbox{$<$}}
{\lower.5ex\hbox{$\sim$}}}}
\newcommand{\myref}[1]{(\ref{#1})}

\newcommand{\ben}{\begin{enumerate}}
\newcommand{\een}{\end{enumerate}}
\newcommand{\bit}{\begin{itemize}}
\newcommand{\eit}{\end{itemize}}
\newcommand{\Cbf}{{\bf C}}

\newcommand{\tmsq}{\tilde m^2}

\newcommand{\mfund}{M_*}

\newcommand{\sqtw}{\sqrt{2}}
\newcommand{\WNR}{W_{\text{NR}}}

\def\[{\left [}
\def\]{\right ]}
\def\({\left (}
\def\){\right )}

\begin{document}

\title{Against Tachyophobia}

\author{John Ellis$^a$}
\email{John.Ellis@cern.ch}

\author{Joel Giedt$^b$}
\email{giedtj@rpi.edu}

\author{Oleg Lebedev$^a$}
\email{Oleg.Lebedev@cern.ch}

\author{Keith Olive$^c$}
\email{olive@physics.umn.edu}

\author{Mark Srednicki$^d$}
\email{mark@physics.ucsb.edu}
\affiliation{
$^a$PH-TH, Case C01600, CERN, CH-1211 Geneva 23, Switzerland \\
$^b$Department of Physics, Applied Physics and Astronomy,
Rensselaer Polytechnic Institute, 110 8th Street, Troy, NY 12180 USA \\
$^c$Fine Theoretical Physics Institute, 116 Church Street Southeast,
University of Minnesota, Minneapolis, 55455 USA \\
$^d$Department of Physics, 3019 Broida Hall, University of California,
Santa Barbara, CA 93106 USA}

\preprint{CERN-PH-TH/2008-135}
\preprint{UMN--TH--2702/08}
\preprint{FTPI--MINN--08/23}

\date{June~23, 2008}

\begin{abstract}
We examine the possible extension of the parameter space of the minimal supersymmetric
extension of the Standard Model (MSSM), as expressed via
the renormalization-group equations in terms of universal soft 
supersymmetry-breaking terms at the unification scale, 
to include tachyonic input scalar masses.
Many models with negative masses-squared for scalars at the unification scale may be viable
because the small sizes of the masses-squared
allow them to change signs during the renormalization-group evolution to the electroweak scale.  
However, in many cases, there is, in addition to the electroweak
vacuum, a much deeper high-scale vacuum located along some F- and 
D-flat direction in the effective potential for the MSSM.
We perform a numerical search for such vacua in both the CMSSM and the NUHM.
We discuss the circumstances under which the existence of such a
deep charge- and color-breaking vacuum is consistent with standard
cosmology. A crucial role is played by the inflation--induced 
scalar masses, whereas  thermal effects are often irrelevant.

\end{abstract}



\maketitle

\section{Motivation}
Understanding the allowed parameter space in versions of the Minimal
Supersymmetric extension of the Standard Model (MSSM) with
Grand Unified Theory (GUT) inspired boundary conditions is a research programme
that gains motivation at the onset of the LHC.  In the constrained MSSM (CMSSM), 
in which gaugino and scalar masses are unified at the GUT scale, and its
generalizations with non-universal boundary conditions, e.g., for the Higgs scalar masses (NUHM),
the identity and relic density of the lightest supersymmetric particle (LSP) place 
strong constraints on the parameter space~\cite{cmssmwmap}.
For example, in models with small values of the unified scalar mass, $m_0$,
and large values of the gaugino mass, $m_{1/2}$, the LSP is typically the 
lighter spartner of the $\tau$ lepton.  This region of the MSSM parameter space 
would predict charged dark matter, and hence would be excluded if $R$-parity is conserved and the
$\tilde \tau_1$ is stable.  However, this corner of parameter
space might be allowed if the gravitino is lighter than the $\tilde \tau_1$, and
becomes the LSP~\cite{gdm}, as may occur in mSUGRA models~\cite{vcmssm}. 
In this case, the $\tilde \tau_1$ decays to the gravitino LSP, and is subject to
important astrophysical constraints that do not exclude this region.  
If the gravitino is the LSP, the CMSSM parameter space 
may even extend to {\em negative} values of $m_0^2$~\cite{m2neg}{\footnote{Negative scalar 
masses-squared also appear at the high-energy scale~\cite{Lebedev:2005ge} in a version of 
mirage mediation~\cite{Choi:2005hd}.}}. In this paper, we argue against excluding all of this
region because of unreasoning tachyophobia.

One can define an effective MSSM model by specifying its mass parameters at the weak scale.
In models with relatively light squark masses, the renormalization-group equations (RGEs) for
the scalar masses may then lead to $m^2(Q_0) = 0$, for some value of $Q_0$ in the range 
$M_W < Q_0 < M_{GUT}$,
with $m^2(Q) < 0$ for $Q > Q_0$.  This raises the question whether there is a dangerous charge- 
and color-breaking (CCB) vacuum~\footnote{For discussions concerning CCB vacua in the
CMSSM, see~\cite{ccb}.}.  The answer to this question depends on two factors: whether
there are potentially large
logarithmic corrections to the potential which are not absorbed in the running 
of mass parameters, and whether there are significant non-renormalizable terms in
the effective potential.
In general these vacua, determined by minimization of the tree--level scalar potential,
occur with vacuum expectation values (vevs) of order $v^2 \sim |m^2| \ll Q_0^2$.
However, $m^2(Q)<0$ and the existence of F- and D-flat directions in the MSSM leads to
runaway to $v(Q) \sim Q$ for $Q > Q_0$, where logarithmic
corrections are small.  Such a CCB vacuum would exist if there were no
non-renormalizable terms in the effective potential, but such terms are in general
present, and their magnitudes determine where the runaway vev is
stabilized and hence whether the existence and location of such a CCB
vacuum can be calculated reliably.  Thus, as argued in~\cite{FORS}, 
for certain parameter choices, vevs will indeed be generated
with $v(Q) \sim Q$ for $Q > Q_0$, ensuring the existence of the CCB
vacuum unless new physics between the GUT and electroweak scale is introduced. 
Here we will enlarge upon this idea and identify the flat directions
that are primarily responsible in the CMSSM and NUHM.

However, it is possible that such CCBs
could be tolerated~\cite{RR}, if the
Universe would have fallen naturally into our false electroweak (EW) vacuum as the cosmological
temperature decreased, and if the lifetime of this vacuum for
tunnelling into the true CCB vacuum is much longer than the present age of the
Universe. Whether the EW vacuum is in fact preferred by cosmology
depends, in particular, on the scalar masses-squared generated during inflation. 
If these masses-squared are
positive and of the order of the square of the Hubble parameter, the `more symmetric' EW
vacuum is favored. On the other hand, if these are negative, as in  the Affleck-Dine scenario
for baryogenesis~\cite{AD},  the Universe would remain trapped in the 
true CCB vacuum~\cite{FORSS}.

The present article delineates regions of
parameter space for which high-scale CCB vacua are present.
We  study  both the CMSSM and less constrained models with
non-universal Higgs masses (NUHM)~\cite{nonu,nuhm}.
For each choice of GUT-scale parameters, we follow the analysis of~\cite{GKM} for determining
the set of problematic flat directions and the lowest order
of GUT- or Planck-scale non-renormalizable operators
whose appearance might lift the flat direction. The existence of CCB vacua will depend
on both the order of the non-renormalizable operator and the fundamental scale
associated with it.  
In the case of the CMSSM, there are wedges of parameter space with $m^2_0 < 0$
where no MSSM sfermion is tachyonic at the EW scale. However, these regions generally
have calculable CCB vacua, assuming that the mass scale $\mfund$ in the non-renormalizable
interaction that stabilizes the high-scale vacuum is greater than or equal to $M_{GUT}$. 
Such regions would be acceptable in suitable
cosmological scenarios that populate exclusively the EW vacuum. 
Similar questionable
tachyonic regions occur also in the NUHM, even if $m^2_0 > 0$,
due to the (independent) squared masses for Higgses being negative.
However, we
emphasize again that these regions would be acceptable in cosmological scenarios
that avoid populating high-scale vacua: one should not be unreasoningly tachyophobic.

\section{CCBs along Flat Directions in the MSSM}
\label{summ}

The tree-level scalar potential can be written schematically
(including soft supersymmetry-breaking contributions) as 
\begin{equation}
V=\frac12 \tmsq(Q)\phi^2 + \frac14 \lambda(Q)\phi^4,
\label{pot}
\end{equation}
where, for simplicity, we have neglected cubic terms,
and $\lambda$ may vanish along some directions in $\phi$ space.
We assume as well that some set of soft supersymmetry-breaking scalar masses
have $\tmsq(Q) < 0$ for $Q$ above some scale $Q_0$, for a particular choice 
of GUT-scale parameters in the CMSSM or NUHM. 
Along a generic direction in field space, with $\lambda(Q) \sim g^4$
where $g$ is some gauge coupling, minimization of the tree-level
potential yields a vev at $\phi = v(Q)=[-\tmsq(Q)/\lambda(Q)]^{1/2}$
which is of similar order to the supersymmetry-breaking scale, $m_{3/2}$. 
However, one-loop corrections to the potential will have the form 
$\Delta V \sim \tmsq(Q) v(Q)^2 \ln [v(Q)^2/Q^2]$ and,
for $v \ll Q_0$, the large logarithms may reduce 
significantly the reliability of the CCB vacuum calculation.

On the other hand, the appearance of
negative mass-squared $\tmsq$ would have important implications for the 
moduli $\eta$ of the flat directions in the MSSM where $\lambda = 0$,
since the tree-level solution yields a runaway vev. However, we 
expect high-scale non-renomalizable operators to regulate the runaway behaviour
in such a case. In the presence of a non-renormalizable superpotential $\WNR$ of degree $n$
(see Appendix A), the effective potential becomes
\beq
V = \tmsq |\eta|^2 + \frac{|u|^2}{\mfund^{2(n-3)}} |\eta|^{2(n-1)}
\label{potnr}
\eeq
and the runaway direction for the modulus $\eta$ is stabilized.
In (\ref{potnr}), $u$ is an $\ord{1}$ dimensionless coupling constant and $\mfund$
is the cutoff scale associated with the dynamics that
generates $\WNR$.  Depending on the circumstance, it could be
between the unification scale and the Planck scale.
The modulus $\eta$ then acquires a vev whose order of magnitude is given
by:
\beq
v \equiv  \[ -\tmsq \mfund^{2(n-3)} \]^{\frac{1}{2(n-2)}}
\quad \( \sim |\eta| \) .
\label{pwr}
\eeq
General properties of flat directions and the operators which lift them
are found in Appendix~A, and examples of specific MSSM flat directions are given in Appendix B.

The one-loop correction
to the scalar potential can be written as
\beq
\Delta V_{\text{1-loop}} = \frac{N}{64\pi^2} g^2 v^2 \tilde m_{\text{avg}}^2
\ln \frac{g^2 v^2}{Q^2} ,
\label{v1l}
\eeq
where $N$ is the number of multiplets
that  get masses at the scale $m=gv$.  The cancellation
between boson and fermion loops has been taken into account,
as is apparent from the factor $\tilde m_{\text{avg}}^2$.
This is an average soft supersymmetry-breaking mass-squared, determined from
the one-loop contributions of the $N$ states.

If $v$ is small, the overall loop correction can be large
relative to the tree-level potential~\myref{potnr}, which
satisfies $V \sim \ord{\tmsq v^2}$ near the 
minimum, rendering the tree-level analysis unreliable.  We introduce
a parameter $\e$ that represents the boundary where this large-logarithm
problem occurs:
\beq
v \lappeq \e Q \Rightarrow \text{unreliable}, 
\quad v \gappeq \e Q \Rightarrow \text{reliable}.
\label{reliability}
\eeq
Since $v$ shuts off at low $Q$ in the models we consider, 
it is possible in principle that the tree-level
analysis may never be reliable.  We have nothing to
say about such models in this article, except that
they are not excluded by our analysis.  On the other
hand, when we find a vev $v$ and there is a regime
of $Q$ for which this tree-level result is reliable,
a vacuum state really exists at some large value of the
modulus $\eta$.  The existence of this deep vacuum
at large field values---generally color or charge breaking (CCB)---has
cosmological implications, which we discuss below.
To determine the parameter $\e$, we note that
the loop correction \myref{v1l} is
comparable to the tree term $\tmsq v^2$ when:
\beq
\frac{N}{64\pi^2} g^2 \ln \frac{v^2}{Q^2} \sim -1
\quad \Rightarrow \quad \e = \exp(-32 \pi^2/N g^2).
\label{epse}
\eeq
When $v \geq \e Q$, we can reliably state that at least two vacua
exist:  the electroweak (good) vacuum and the high-scale (bad) one.

Now a word on numbers:  in \myref{epse}, we should
take $1/10 < g^2 < 1/2$, corresponding to the running
constants between the EW and GUT scales.  The number
$N$ depends on the flat direction, but would typically
range between $\ord{10}$ and $\ord{100}$.  For the
larger value of $N$, we have $\e \lappeq \ord{10^{-3}}$; moreover, $\e$ falls
exponentially as $N$ decreases, so it is generically much smaller
than $10^{-3}$. In other words, the loop
factor suppression $g^2 / 64\pi^2$ means that the
logarithmic enhancement must be quite large before
the tree-level analysis becomes unreliable.  
As a consequence, the RG-improved tree-level analysis is
generally a reliable indicator of the high-scale vacuum.

As an example, consider the $LLe^c$
flat directions in the MSSM. Eleven chiral multiplets participate
in the mass matrix once renormalizable
superpotential terms are accounted
for, namely  $e_i, \nu_i, e_i^c, H_d^0, H_d^-$.
The moduli space is $\Cbf^3$ once D- and F-flatness
constraints are taken into account \cite{GKM}.
Thus eight chiral multiplets get $\ord{v}$ masses,
although in some cases they are  suppressed by very
small Yukawa couplings $y_i^e$, corresponding to
D-moduli whose flatness is lifted by the renormalizable 
superpotential~\footnote{The $LLe$ moduli that inhabit $\Cbf^3$ are lifted
by the non-renormalizable superpotential and
get $\ord{\tilde m}$ masses, which can be ignored
in the loop corrections.}.
Furthermore, $SU(2) \times U(1)$ gauge multiplets 
get $\ord{gv}$ masses.  Each chiral multiplet
contributes 2 to $N$, because it includes a complex
scalar and a Weyl fermion.  A similar contribution
comes from each of the four vector multiplets
contained in $SU(2) \times U(1)$.  Thus $N=2(8+4)=24$.
Even if we assume the GUT-scale value $g^2 = 1/2$, we already get
$\e \sim 4 \times 10^{-12}$.  In actuality, for
many of the masses we should use $(y_i^e)^2 \ll g^2$
and $\e$ is even much smaller still.  The result is that
we trust the tree-level analysis for
practically all $v/Q$: it is a robust result that we have at least these two vacua,
and the cosmological arguments apply.

\begin{figure}
\begin{center}
\includegraphics[width=5in,height=4in]{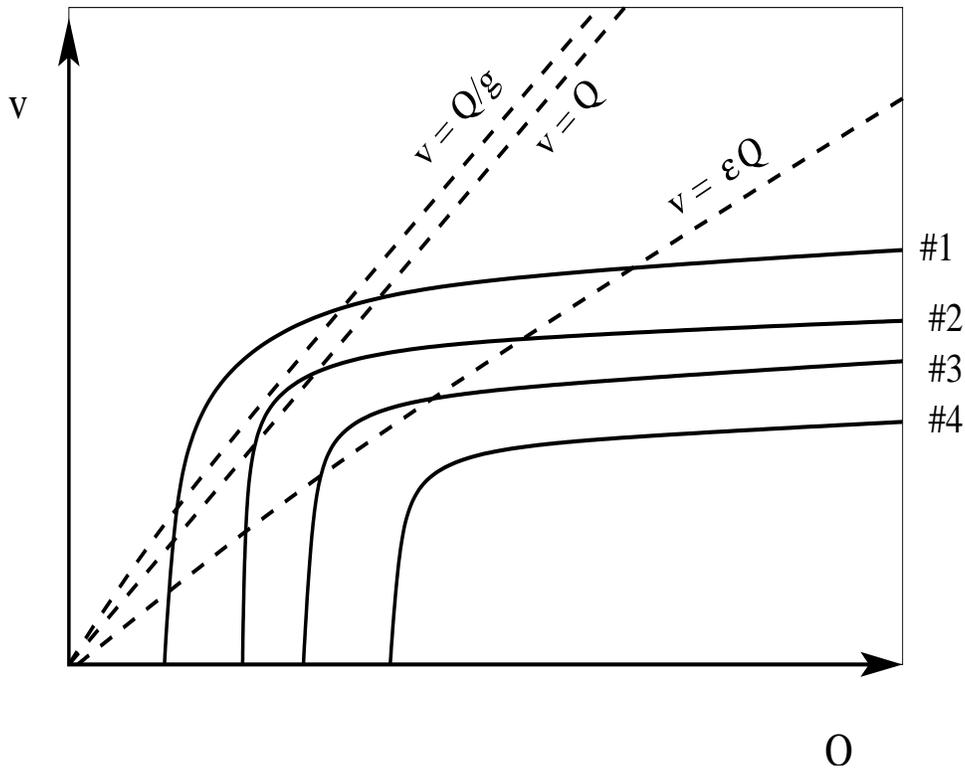}
\caption{\it Curves of the flat direction vev $v=v(Q)$ (solid)
versus various constraint curves (dashed). In regions where $v(Q)$ lies above
the dashed lines, the tree-level effective potential is a reliable indicator for the
existence of a high-scale (bad) vacuum, according to the criterion (\protect\ref{reliability}).
\label{vevQ}
}
\end{center}
\end{figure}

Fig.~\ref{vevQ} represents schematically several generic cases for the possible
variation with $Q$ of the vev $v(Q)$. For each of the curves \#1 -- \#4, $v(Q) \ne 0$
at large $Q$, but $v(Q) \to 0$ at small $Q$. In each of the cases \#1 -- \#3, the
condition (\ref{reliability}) is satisfied over some range of $Q$ and hence,
according to the arguments presented above,
the existence of a high-scale vacuum can be predicted reliably in each of these cases.
These are representative of what happens
in much of the parameter space that we study below in specific CMSSM and NUHM scenarios.
On the other hand, when we consider curve \#4 in Fig.~\ref{vevQ}, we see that, over the
entire range of $Q$, the tree-level prediction cannot be trusted, 
owing to the persistence of large logarithms.  For this reason, we
exercise caution and choose  not  to apply
the cosmological constraint in such a case.
In some specific cases, a one- or two-loop analysis
might be reliable, and the possibility of
a high-scale (bad) minimum could be examined in more detail.  
However, studies of such possibilities lie beyond the scope of this work,
where we restrict ourselves conservatively to the criterion (\ref{reliability}).

\section{The Cosmological Constraint}
\label{outl}

Having established the criteria for determining the existence of a 
high-scale vacuum (which we apply to specific CMSSM and NUHM models 
in the following two sections),
we now turn our attention to the cosmology of models with high-scale vacua,
asking how problematic their existence may be.
Such a global CCB vacuum with $v \sim Q \gg M_W$ is separated from the local 
charge- and color-conserving minimum at the origin (or the EW scale if the Higgses are
the scalars in question) by a potential barrier.  As argued in~\cite{RR},
this model remains perfectly acceptable if the Universe is trapped in 
the local minimum near or at the origin, as the timescale for
producing a bubble of lower high-scale vacuum is generally much longer than the age of the Universe.

However, in models with negative 
$m_0^2 (M_{\rm GUT})$, there are many such vacua corresponding to different flat directions,
and each has a larger domain of attraction than that of  the EW 
vacuum. Thus, for arbitrary initial conditions, the system is overwhelmingly
likely to end up in one of the `bad' vacua. 
However, this question must be analyzed in some appropriate cosmological setting, 
which introduces two extra ingredients  affecting  the shape of the effective potential:
inflation and thermal effects. 
During inflation, supersymmetry is broken 
by the vacuum energy, which results in an extra contribution
to the soft scalar masses, of the general form~\cite{drt}
\begin{equation}
\Delta m^2 =c  H^2 \;,
\end{equation}
where $H$ is the Hubble expansion rate during the inflationary epoch. 
The dynamics of the flat directions depend crucially on the sign  of $c$, which is model-dependent.
For a minimal K\"ahler potential,  $c=3$. On the other hand,  for a no-scale  K\"ahler potential, 
the induced scalar masses are zero at the tree level and 
loop corrections generate $c<0$ for flat directions not involving the stop~\cite{gmo}.

We now consider different possibilities for the coefficient $c$,
considering first the possibility $c>0$. We assume also that the initial value of the flat-direction
field $\phi$ is $ 10^{17}$~GeV, in which case $V(\phi) \ll V_{\rm inflaton}$, and the evolution of $\phi$ 
does not affect the Hubble constant  significantly until the late stage when $H \sim
m_{3/2} $. In this case, $\phi$ obeys the equation of motion
\begin{equation}
\ddot{\phi} + 3H \dot{\phi} + V'(\phi)=0 \;,
\label{evolution}
\end{equation}
where $H$ is a slowly-varying function of time that can be described by an adiabatic approximation.
For our purposes, $V'$ can be approximated by $cH^2 \phi$ until $H$ becomes
comparable to  $m_{3/2}$. The effect of the  non--renormalizable term 
in $V(\phi)$ is less significant. The general solution of (\ref{evolution}) is
\begin{equation}
\phi(t)= C_1 \exp \left[ {1\over 2} H (-3 + \sqrt{9-4c } )~t \right] +
         C_2 \exp \left[  {1\over 2} H (-3 - \sqrt{9-4c } )~t \right] \;,
\end{equation}
where $C_{1,2}$ are determined by the initial conditions.

We consider first the  $classical$ evolution. 
For $c= {\cal O}(1)$,  the magnitude of $\phi$ scales as 
$\phi(t) \sim \exp [- {\cal O}(1) Ht ]$. Thus, within 5-10 Hubble times, $\phi$
will be of order $H$ even if its initial value was very large~\footnote{Here we
take $H\sim 10^{13-14}$ GeV and $\phi_0 \sim 10^{17}$ GeV.}
and, within the next $\sim$30 Hubble times, $\phi$ will be of the electroweak scale.

We next take into account the de Sitter quantum fluctuations,
which play an important role in the dynamics of flat directions~\cite{FORSS}.
For a scalar field of mass $\sim H$ \cite{lin2},
\begin{equation} 
\langle \phi^2 \rangle \sim H^2. 
\end{equation}
It is then clear that at large $\phi$, the classical damping dominates and when
$\phi$ becomes comparable to $H$, these quantum fluctuations are as important as the 
classical evolution. From that moment on, $\phi$ undergoes random oscillations
of order $H$ per Hubble time  plus classical damping which decreases its magnitude
by a factor ${\cal O}(1)$ per Hubble time.  
As $H(t)$ decreases, so does the amplitude 
of random oscillations about the origin. At the time when the soft  term 
$m^2_0$  
becomes relevant (which is after the `inflaton oscillation' era), 
the amplitude of oscillations is  ${\cal O}(m_{3/2})$.
The field settles at the origin and the quantum oscillations are too small
to reach the barrier separating the two minima,  which is further  than 
${\cal O}(m_{3/2})$ away from the origin. Therefore, 
the presence of a deep CCB minimum would not be problematic 
for $c \sim 1$.

However, these conclusions do not in general apply for small positive $c$: $0< c \ll 1$,
which is a borderline case. First, the classical evolution
is slow and, secondly, the amplitude of quantum oscillations is larger: 
$ \langle \phi^2 \rangle \simeq {3H^2 / (8\pi^2 c )}.$  Whether the field
settles at the origin at the end of inflation  depends on further specifics
of the inflationary model as well as the magnitude of $c$.

For $c\leq 0$ (see~\cite{drt}),   
the minimum of the potential during (as well as after)  inflation is at large $\phi$.
Classical evolution will drive $\phi$ towards this minimum, whose  position
is a slowly-evolving function of time. At $H \sim {m_{3/2}}$, 
the field  freezes at the CCB minimum, and 
the quantum fluctuations do not play any significant role.
It is important to note also that  thermal effects are irrelevant at large $\phi$
and cannot destabilize the CCB minimum. This is because
all the fields $\phi$ couple and receive masses of order $\phi$ (multiplied
by the gauge or Yukawa couplings). Thus they
are heavy and cannot be thermalized. Consequently, no thermal mass term 
$T^2 \phi^2$ is generated by thermal loops~\footnote{Note that, although
$\phi$ itself is light, it has no self--interactions at the renormalizable level.}.
 
We thus conclude that, for $c\leq 0$,   
the presence of deep  CCB minima is ruled out by
cosmological considerations. This applies in particular to models with the 
Heisenberg symmetry~\cite{BG}, including no-scale models of
supergravity~\cite{ns}. Furthermore, this precludes the possibility
of the Affleck-Dine (AD) mechanism for baryogenesis~\cite{AD}, which
requires negative $c$~\cite{drt}. 
Concretely, this applies to the cases \#1 -- \#3 of Fig.~\ref{vevQ}, in which the
existence of such a bad high-scale CCB can be predicted reliably. On the
other hand, even these cases would not be excluded for positive $c \sim 1$,
and possibly also for small positive $c$.

We remark finally that we have
neglected  the trilinear soft supersymmetry-breaking $A$-terms in the above considerations. 
In general, one expects $A$-terms of order $H$
to be generated during inflation. If their magnitude is a few times larger than 
$H$, an additional local minimum at large $\phi$ appears even for positive 
$c\sim 1$. Given a large initial value of $\phi$, the field will evolve
to this minimum and remain there after inflation. Thus, deep CCB vacua can be problematic  
even for $c \sim 1$, if the $A$-terms are sufficiently large.

In what follows, we will examine specific CMSSM and NUHM parameter sets 
for which $\tmsq < 0$ at some high renormalization scale, with the aim of
elucidating whether one should worry about them.

\section{Tachyons in the CMSSM}
We begin by considering the parameter space of
the CMSSM, in which all gaugino masses are unified at a common scale 
(where gauge-coupling unification occurs) with the common value $m_{1/2}$.  Similarly,
all soft scalar masses are unified at the same renormalization scale with a common value $m_0$,
as are the trilinear terms with a common value $A_0$.  The remaining free parameters are 
the ratio of the two Higgs vevs, parametrized by $\tan \beta$, and the sign of the $\mu$ parameter.  

\begin{figure}
\begin{center}
\begin{tabular}{cc}
\includegraphics[width=3in]{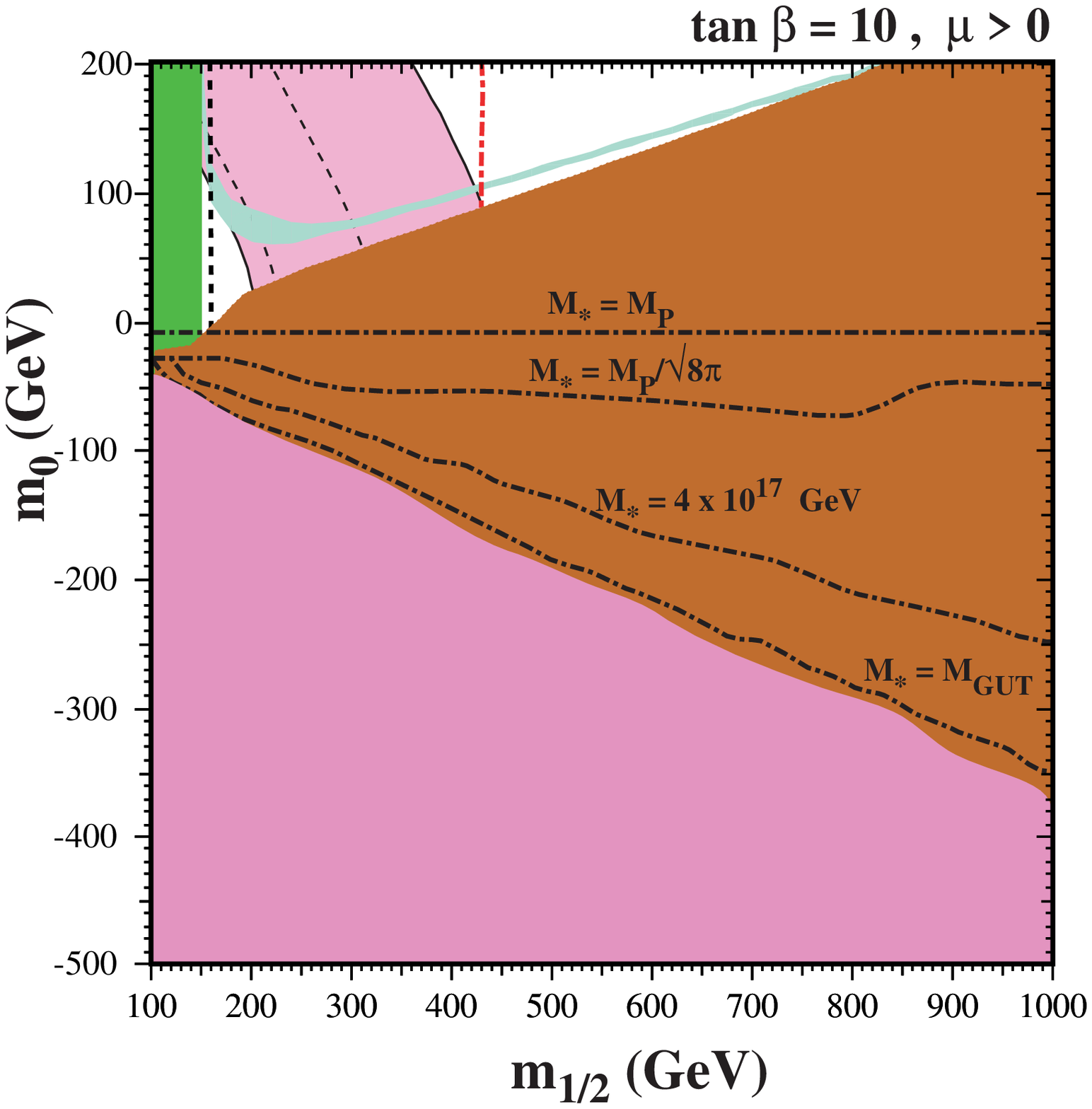} &
\includegraphics[width=3.05in]{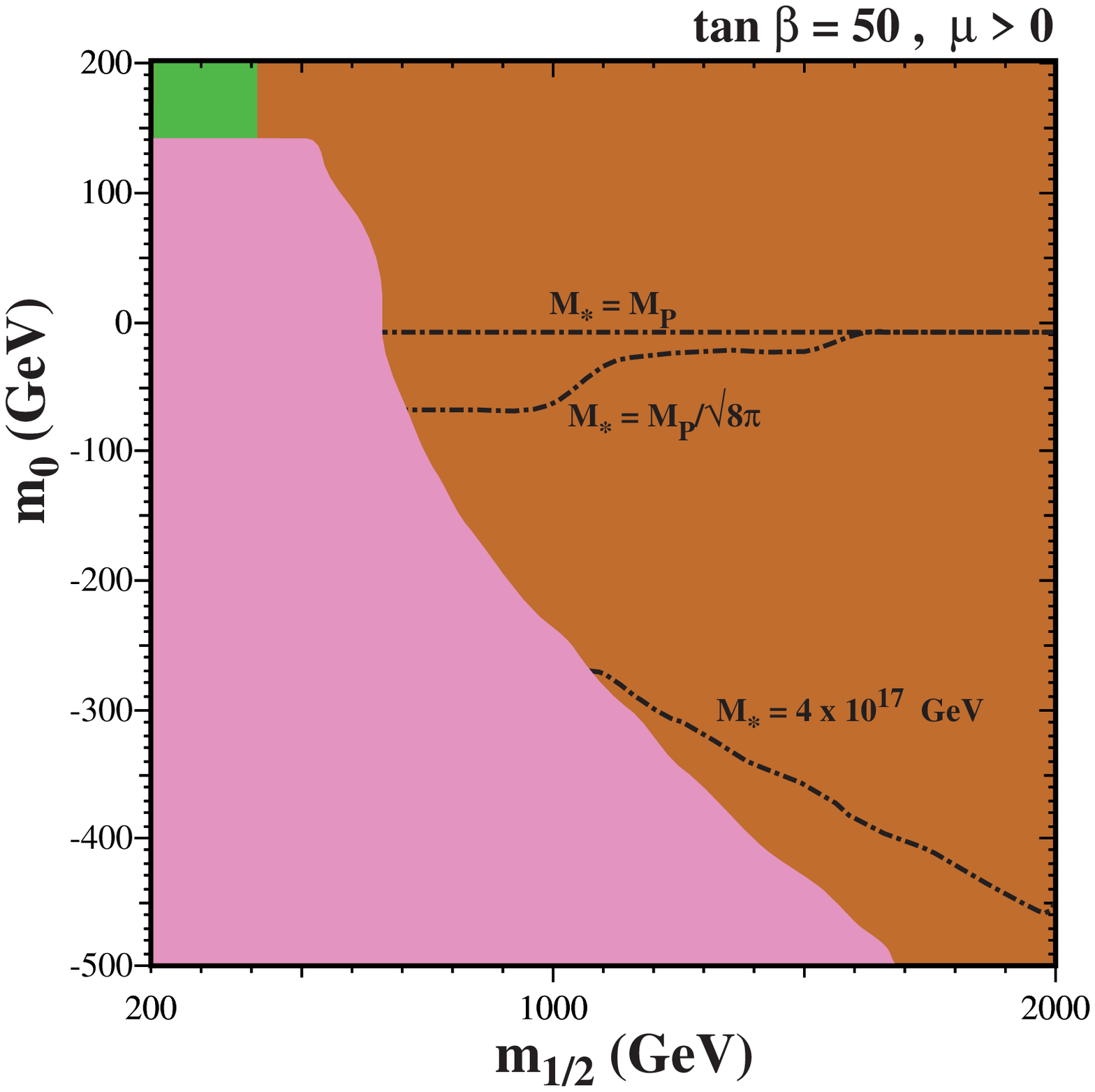} \\
(a) & (b)
\end{tabular}
\caption{\it Portions of the CMSSM $(m_{1/2}, m_0)$ planes for (a) $\tan \beta = 10$
and (b) $\tan \beta = 50$.  The notations are described in the text.  The lower pink
region is excluded because the lighter stau would be tachyonic at the EW scale.
The regions below the dash-dotted curves have problematic high-scale tachyons
for the indicated values of $\mfund$ if $\epsilon = 1$,
while for $\e \lappeq 10^{-3}$ all regions with $m_0^2 < 0$ are excluded.
\label{cmssm}
}
\end{center}
\end{figure}

For each choice of these four input parameters (plus the sign of $\mu$),
the low-energy spectrum can be determined and compared with phenomenological 
and cosmological constraints.  In Fig.~\ref{cmssm}, we show the $(m_{1/2}, m_0)$
plane for two fixed values of $\tan \beta$, both for $\mu > 0$ and $A_0 = 0$.
The sign displayed on the vertical axis is actually the sign of $m_0^2$, 
so what is displayed is $m_0^2/|m_0|$, strictly speaking.
In panel (a), we have fixed $\tan \beta = 10$ and, for $m_0^2 > 0$, we
see results common in many CMSSM studies~\cite{cmssmwmap}.
The dark (brown) shaded region corresponds to the region for which the stau is the LSP
and as such this region is normally excluded unless the gravitino is in fact the LSP~\footnote{Note that,
even in this case, much of the region shown is excluded due to effects during and after
Big-Bang nucleosynthesis involving the bound state of the stau and He: see~\cite{cefos} and references therein.}. The medium (green) shaded region, at low $m_{1/2}$ is excluded by the constraint
arising from the branching ratio of $b \to s \gamma$. The vertical dashed line is the 
chargino mass contour at 104~GeV, and the nearly vertical dot-dashed line is
the Higgs mass contour at 114~GeV, as obtained using FeynHiggs~\cite{FeynHiggs}.
Only regions to the right of these lines are allowed by LEP.
The pink band bordered by black solid curves is the region where supersymmetric
corrections to the Standard-Model calculation of $(g-2)_\mu$ match the experimental measurement 
of $(g-2)_\mu$ within 2-$\sigma$ uncertainties (between the dashed curves agreement occurs at the 
1-$\sigma$ level).  Finally, in the (turquoise) shaded region that tracks the stau LSP boundary
at large $m_{1/2}$, the relic density of the lightest neutralino
would lie in the range of the cold dark matter
density determined by WMAP and other observations~\cite{wmap}, if this neutralino were the LSP.
  
Also shown in Fig.~\ref{cmssm} (a) is a large (pink) shaded region at low $m_{1/2}$
and negative $m_0^2$ where one of the sfermions is tachyonic at the electroweak scale.
This region is also excluded.  Of particular interest to us here is the region where $m_0^2 < 0$
but the shading indicates that the lighter stau is the lightest spartner of a Standard Model
particle but is not tachyonic at the electroweak scale: we repeat that parts of this region may in 
principle be viable~\cite{m2neg} if the gravitino is in fact the LSP.
We perform our scan of the parameter space in this region looking for flat directions which could 
lead to a bad high-scale minimum of the potential.

According to the analysis of \cite{GKM},
the $QQuue$ flat direction is not fully
lifted until degree $n=9$, so from Eq.~\myref{pwr}
we see that $v$ along this direction is quite close to the non-renormalizable
mass scale $\mfund$.  
The dot-dashed curves in Fig.~\ref{cmssm} labeled by $\mfund$ demarcate regions 
for which we find solutions to $v(Q) = Q$: for a given value of $\mfund$, all regions below the 
corresponding curve 
admit solutions to $v(Q) = Q$. Thus, below the curves we find a bad CCB vacuum, that might
invalidate the corresponding choice of parameters, subject to 
the cosmological considerations discussed in the previous Section. 
When $\mfund = M_P$, all regions with $m_0^2 < 0$ have this problem.
Because all of the scalars have $\tmsq_i < 0$ at high enough scale when
$m_0^2<0$, this flat direction always has a reliable bad CCB minimum when we take
$\e \lappeq 10^{-3}$, for all $\mfund \ge M_{GUT}$.  Thus all CMSSM models with $m_0^2<0$ and
$\e \lappeq 10^{-3}$ are problematic according to our analysis.

In panel (b) of Figure \ref{cmssm}, we show the analogous parameter plane when $\tan \beta = 50$,
$\mu > 0$ and $A_0 = 0$.
The part of the $(m_{1/2}, m_0)$ plane shown is dominated by regions where the stau is the
lightest sparticle apart from the gravitino (shaded dark brown) and where the stau is tachyonic
at the EW scale (shaded pink).
The standard CMSSM phenomenologically and  cosmologically acceptable regions occur
at $m_{1/2} > 200$~GeV, and so are not visible in the part of the $(m_{1/2}, m_0)$
plane displayed. The wedge-shaped brown region with 
$m_0^2 \le 0$ has a calculable high-scale CCB minimum, as was the case $\tan \beta = 10$. 
For this reason, we do not display planes with
intermediate choices of $\tan \beta$.
We have also scanned the CMSSM planes with $A_0 \ne 0$, and the results are 
qualitatively similar to those shown here.  

To summarize, because in practice $\e \lappeq 10^{-3}$ for any possible
flat direction, and more especially for the $QQuue$ one, the CMSSM
with $m_0^2<0$ always has a bad CCB vacuum. However, this may not be
populated, when cosmological considerations are taken into account.

\section{Tachyons in the NUHM}
\label{tnu}

One alternative to the CMSSM is the NUHM, where the scalar partners of the quarks and
leptons still unify at $M_{GUT}$, but the soft supersymmetry-breaking scalar masses associated with the 
two Higgs doublets do not. This class of models has, effectively, two additional free parameters
relative to the CMSSM. These are often chosen to be the weak-scale values of 
$\mu$ and the Higgs pseudoscalar mass, $m_A$.  Whilst it is certainly possible within the context of
the NUHM to choose $m_0^2 < 0$ (leading to the same difficulty
with the $QQuue$ flat direction as in the CMSSM), 
problems with CCB can already occur for certain choices
of $\mu$ and $m_A$ even when $m_0^2 >0$.  This is because, when the weak scale $|\mu|$ is 
large, typically one or both of
the Higgs squared masses $m_{1,2}^2=\mu^2+\tmsq_{1,2}$ is negative at the GUT scale, a
problem that is accentuated at small $m_A$.  In this case, 
the squark and slepton masses-squared remain positive throughout the RGE evolution,
avoiding the $QQuue$ flat direction problem, and vevs $v \ll Q$ for all $Q$
may develop along the $H_1 H_2$ flat direction, because this is lifted by a
lower-order non-renormalizable term ($n=4$).
As a consequence, a larger fraction of the parameter space is allowed.

At scales $Q$ not too far above the EW scale, one begins to see the
EW vacuum.  The negative masses-squared of MSSM 
Higgses can lead to unacceptable vacua with CCB also
at this low scale.  We do not include the details of such low-scale 
CCB vacua in the analysis, as this would require a much
more careful treatment of loop corrections and contributions
from all soft supersymmetry-breaking operators.  For this reason, we terminate the
NUHM analysis at $Q=10$ TeV in our numerical studies.

\begin{figure}
\begin{center}
\begin{tabular}{cc}
\includegraphics[width=2.7in]{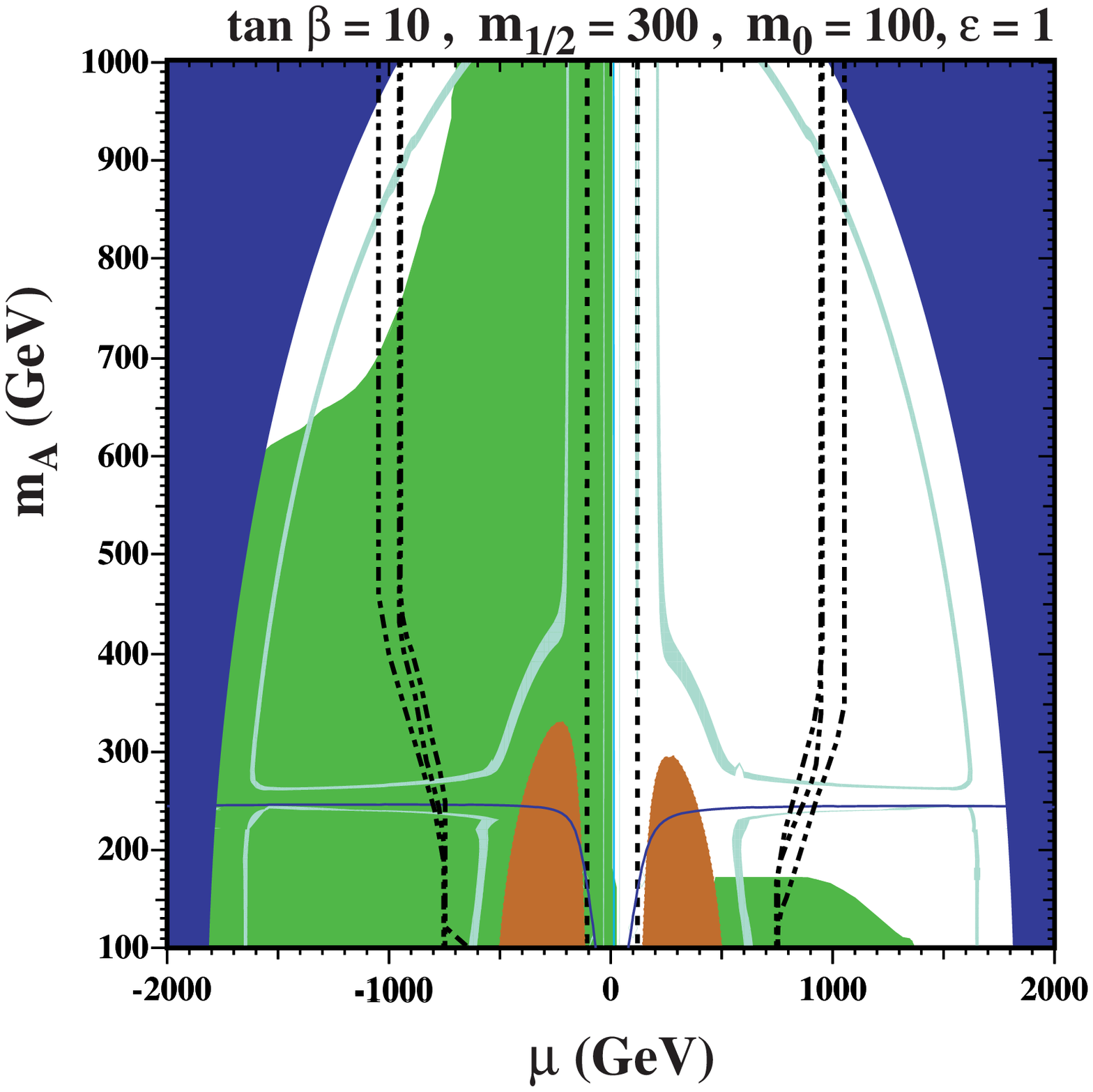} 
& \includegraphics[width=2.7in]{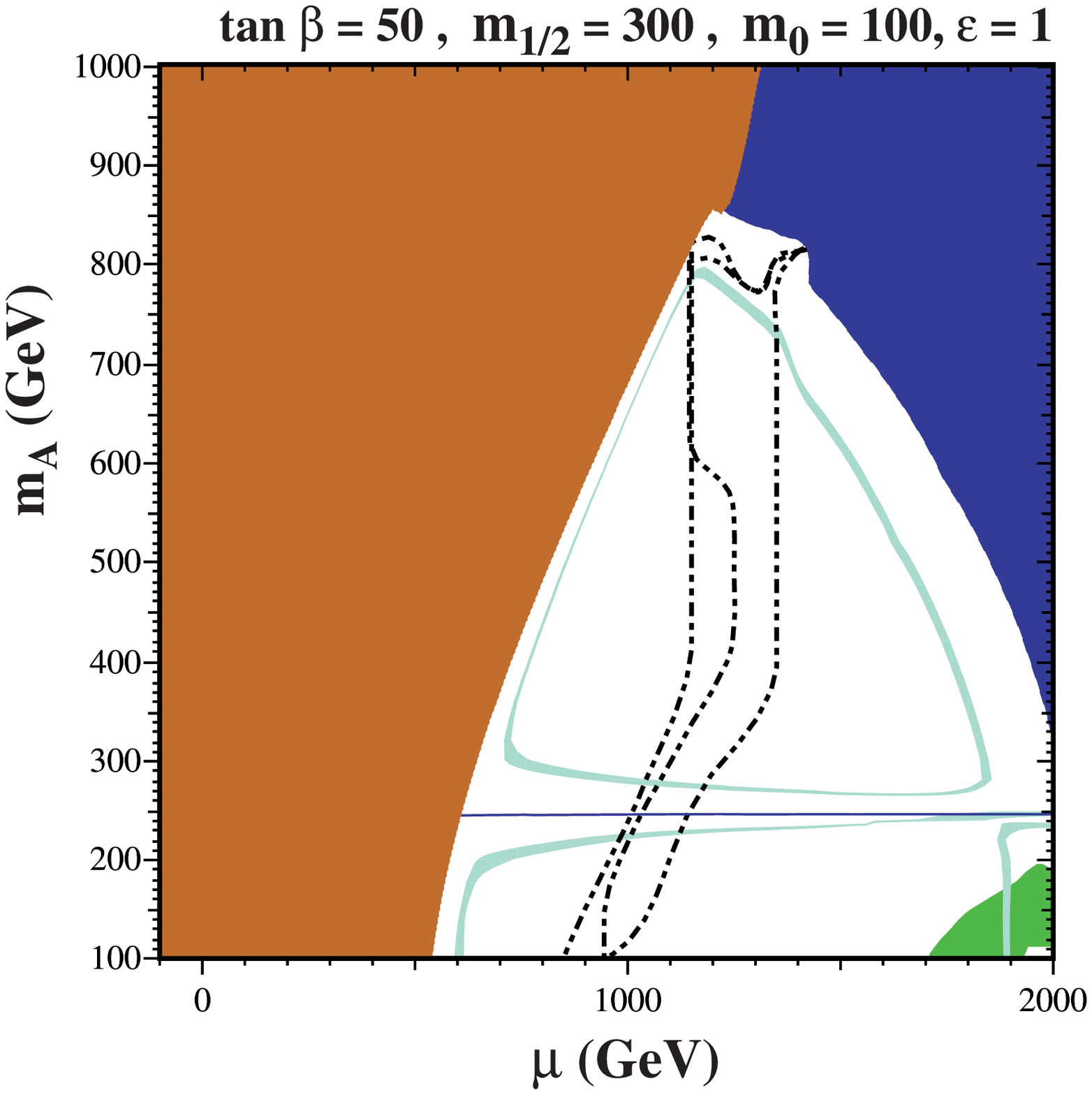} \\
\includegraphics[width=2.7in]{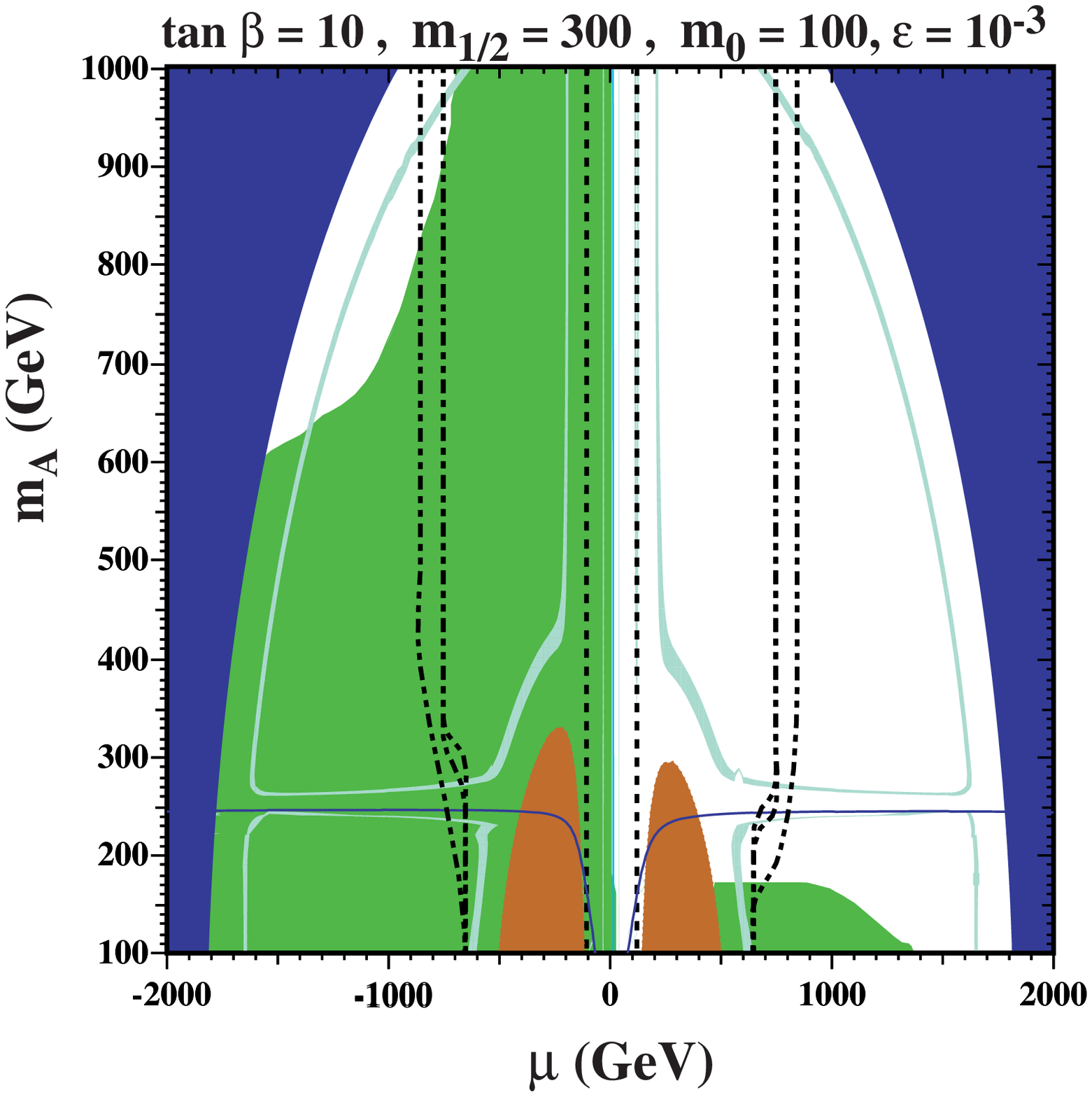} 
& \includegraphics[width=2.7in]{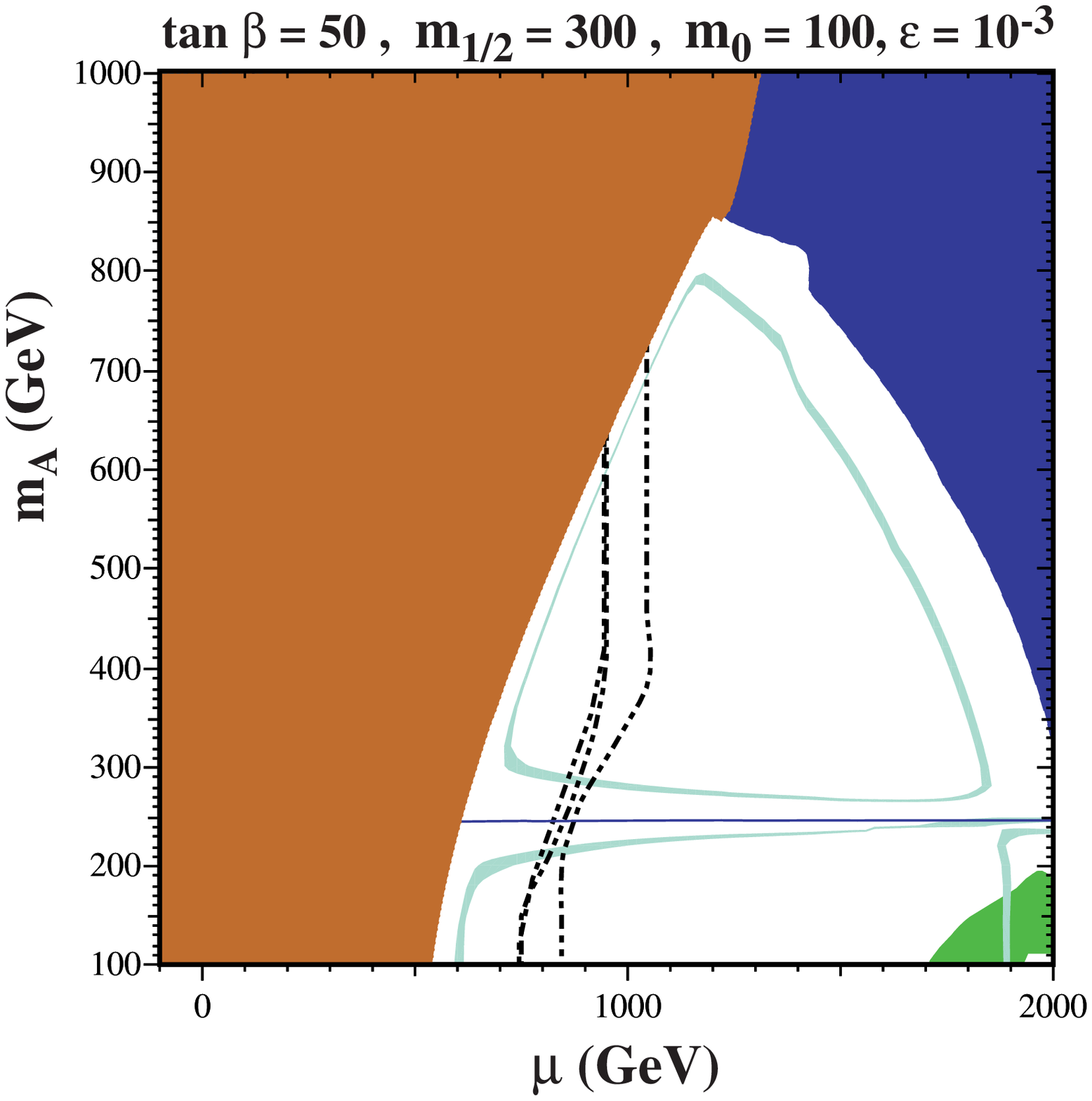} \\
\includegraphics[width=2.7in]{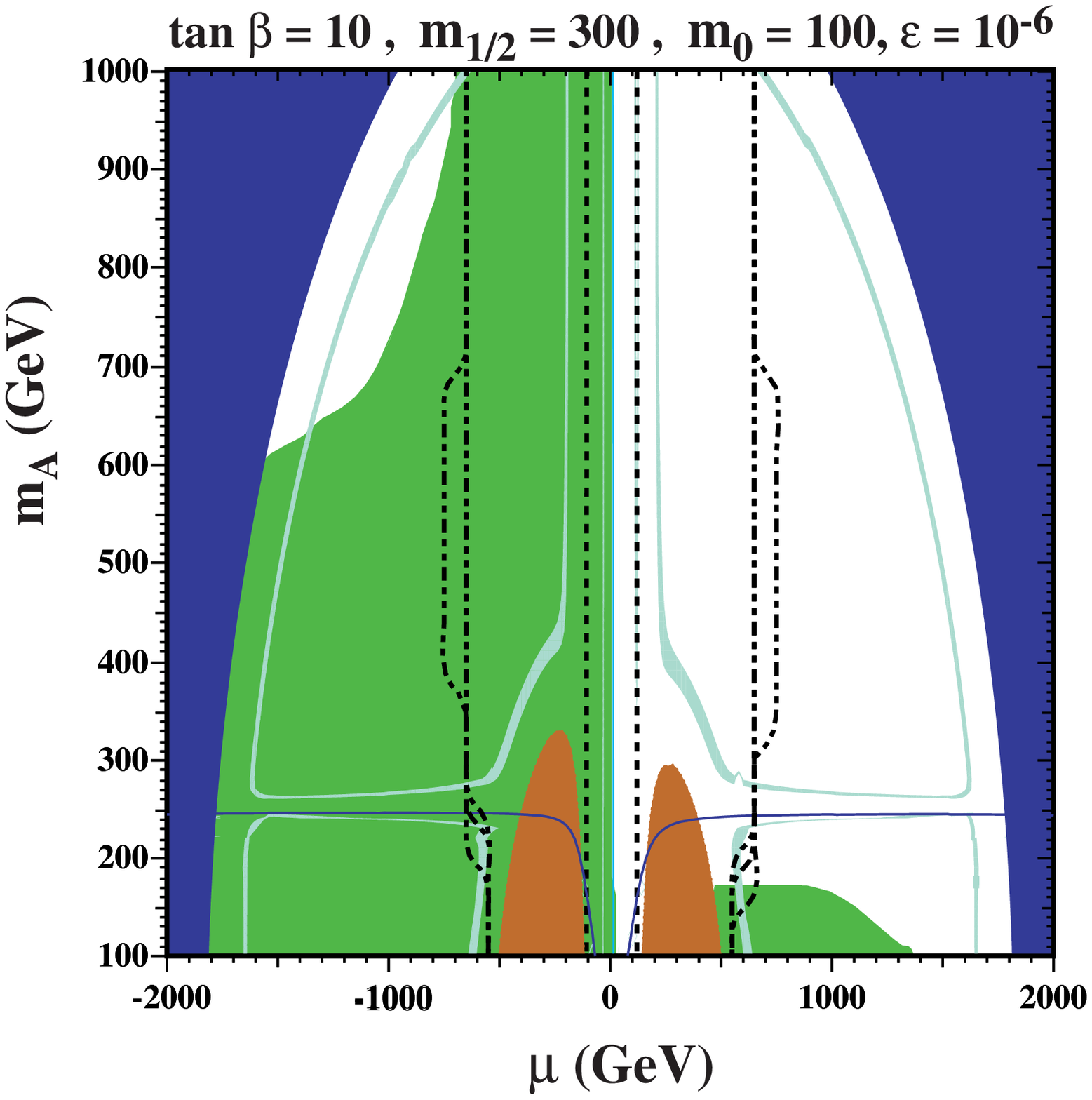} 
& \includegraphics[width=2.7in]{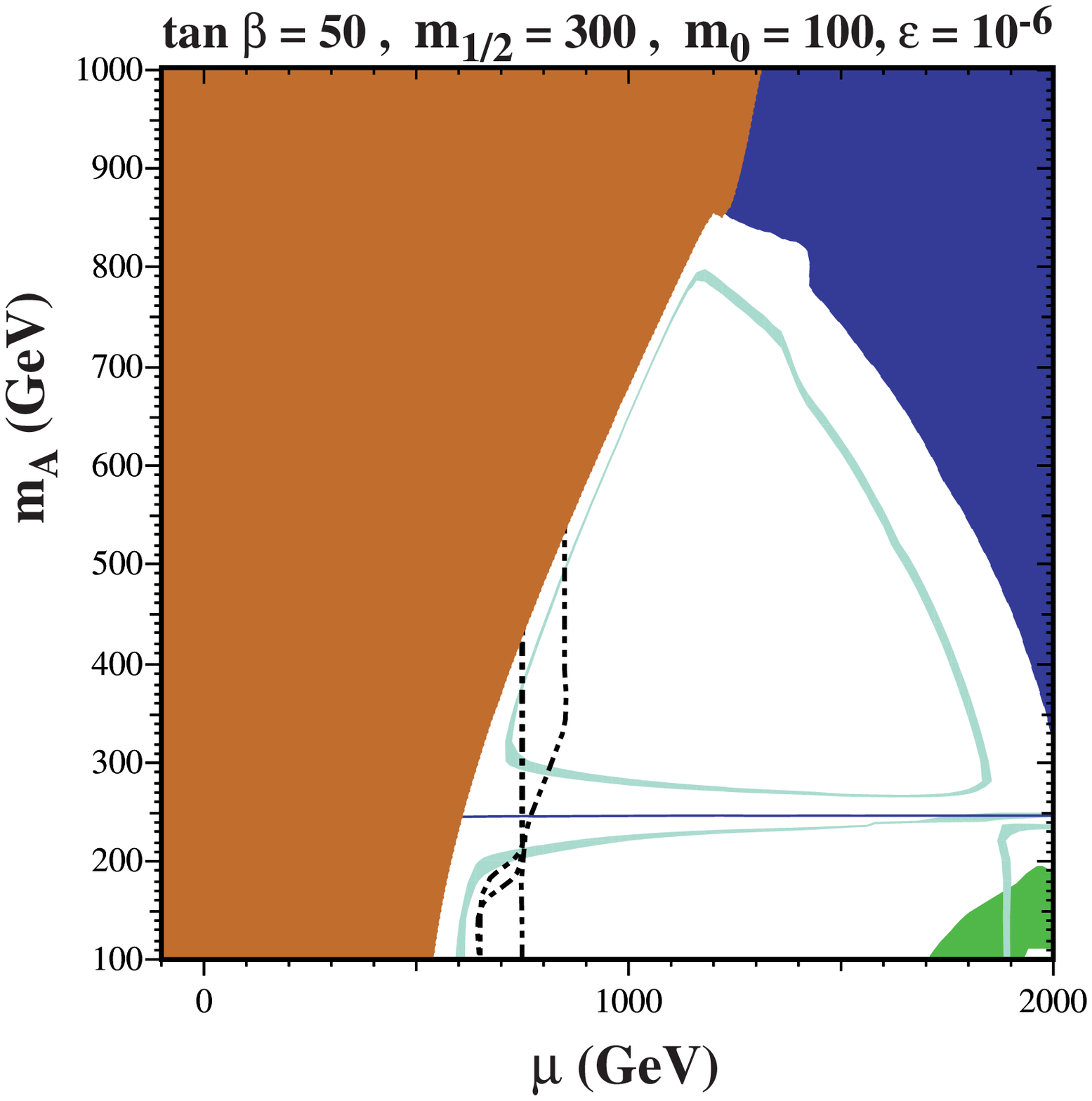} \\
\end{tabular}
\caption{\it Portions of the NUHM $(\mu, m_A)$ planes for for  $m_{1/2} = 300$~GeV, 
$m_0 = 100$~GeV,
$A_0 =0$ and (a, c, e) $\tan \beta = 10$, (b, d, f) $\tan \beta = 50$. The notations are 
described in the text. The regions outside the inner dash-dotted curves have problematic 
high-scale tachyons for higher values of $\mfund$.
\label{vef1}
}
\end{center}
\end{figure}

In Fig.~\ref{vef1}, we show examples of $(\mu, m_A)$ planes in the NUHM
for fixed $m_{1/2} = 300$ GeV, $m_0 = 100$ GeV, and $A_0 = 0$
with $\tan \beta = 10$ (panels a, c and e) and $\tan \beta = 50$ (panels b, d and f).
We use the same shadings as used for the CMSSM to denote regions excluded by
$b \to s \gamma$ (shaded medium (green)),
which excludes much of the parameter space at $\mu < 0$, and a region in which
the lighter stau would be the lightest spartner of a Standard Model particle (shaded dark (brown)).
This includes two areas with relatively
small $\mu$ and $m_A$, for $\tan \beta = 10$, and  most of the left side of the plane. for
$\tan \beta = 50$.
New to this figure are regions shaded dark (blue) for which the sneutrino is the lightest spartner
of a Standard Model particle, 
that are seen at large $\mu$ and $m_A$, for $\tan \beta = 10$, and in the upper right of the planes,
for $\tan \beta = 50$. 
At least parts of the regions with a light stau or sneutrino could be allowed if the gravitino is the LSP.
Once again, the dashed vertical lines at small $\mu$ show the 104~GeV chargino mass
contour.  The thin blue lines show the contour where
$m_A = 2 m_\chi$ and the regions of good neutralino relic density
near these lines correspond to the rapid-annihilation
funnel region familiar from the CMSSM at large $\tan \beta$. 
Other strips with an acceptable neutralino LSP relic density appear in the stau and sneutrino
coannihilation regions, running parallel to boundaries of the brown and blue regions,
and in a `crossover' strip close to the chargino exclusion, where the neutralino is
 a mixed gaugino/higgsino state.
 
We show in each panel of Fig.~\ref{vef1} three dot-dashed contours with differing $\mfund$. 
The inner curves (with the lowest values of $|\mu|$)
correspond to $\mfund = M_P$, whereas the middle and outer curves correspond to
$\mfund = M_P/\sqrt{8 \pi}$ and $M_{GUT}$, respectively. The areas outside 
these contours may be problematic, depending on the cosmological scenario,
as discussed previously. In panel (a) of Fig.~\ref{vef1}, we have chosen $\epsilon = 1$
and $\tan \beta = 10$. The problematic region is when $|\mu| \gappeq 1000$~GeV, reducing to 
$|\mu| \gappeq 750$~GeV for small $m_A$. When $\epsilon$ is decreased to
$10^{-3}$ (panel c) and $10^{-6}$ (panel e), the problematic regions extend
down to smaller values of $|\mu|$, reaching as low as $\sim 550$~GeV for
$\epsilon = 10^{-6}$ and small $m_A$, essentially independent of $\mfund$.

The most immediately noticeable features of the $(\mu, m_A)$
planes for $\tan \beta = 50$, shown in panels (b), (d) and (f)
of Fig.~\ref{vef1}, are the greater extent of the stau LSP region
when $\mu \lappeq 1000$~GeV, and the the greater extent of 
the sneutrino LSP region when $\mu \gappeq 1000$~GeV.
In between, the problematic tachyonic regions depend more
sensitively on the value of $\mfund$ than was the case for
$\tan \beta = 10$, and also vary more as $\epsilon$ is reduced.
Once again, it is the regions of large $\mu$ that are problematic.

Whereas in the CMSSM case the stau was always lighter than the
lightest neutralino in the tachyonic region, so that it could be
cosmologically acceptable only if the gravitino were the LSP,
in the NUHM case the tachyonic region also includes parts of the WMAP strips where
the LSP is the neutralino and it has an acceptable relic density.
The crossover strip, the stau coannihilation region and parts of the 
snu coannihilation strip and parts of the rapid-anihilation
funnel with an acceptable neutralino relic density are all in the non-tachyonic
parts of the $(\mu, m_A)$ planes for $\tan \beta = 10$. These regions also have
acceptable $b \to s \gamma$ for $\mu > 0$. In the case of $\tan \beta = 50$,
parts of the stau coannihilation strip and the rapid-annihilation funnel are
again tachyon-free. We emphasize yet again that the tachyonic regions at
larger $|\mu|$ are not necessarily excluded: that would depend on the
cosmological scenario and whether it avoids the high-scale vacuum in
the early Universe.

\begin{figure}
\begin{center}
\begin{tabular}{cc}
\includegraphics[width=2.7in,height=2.7in]{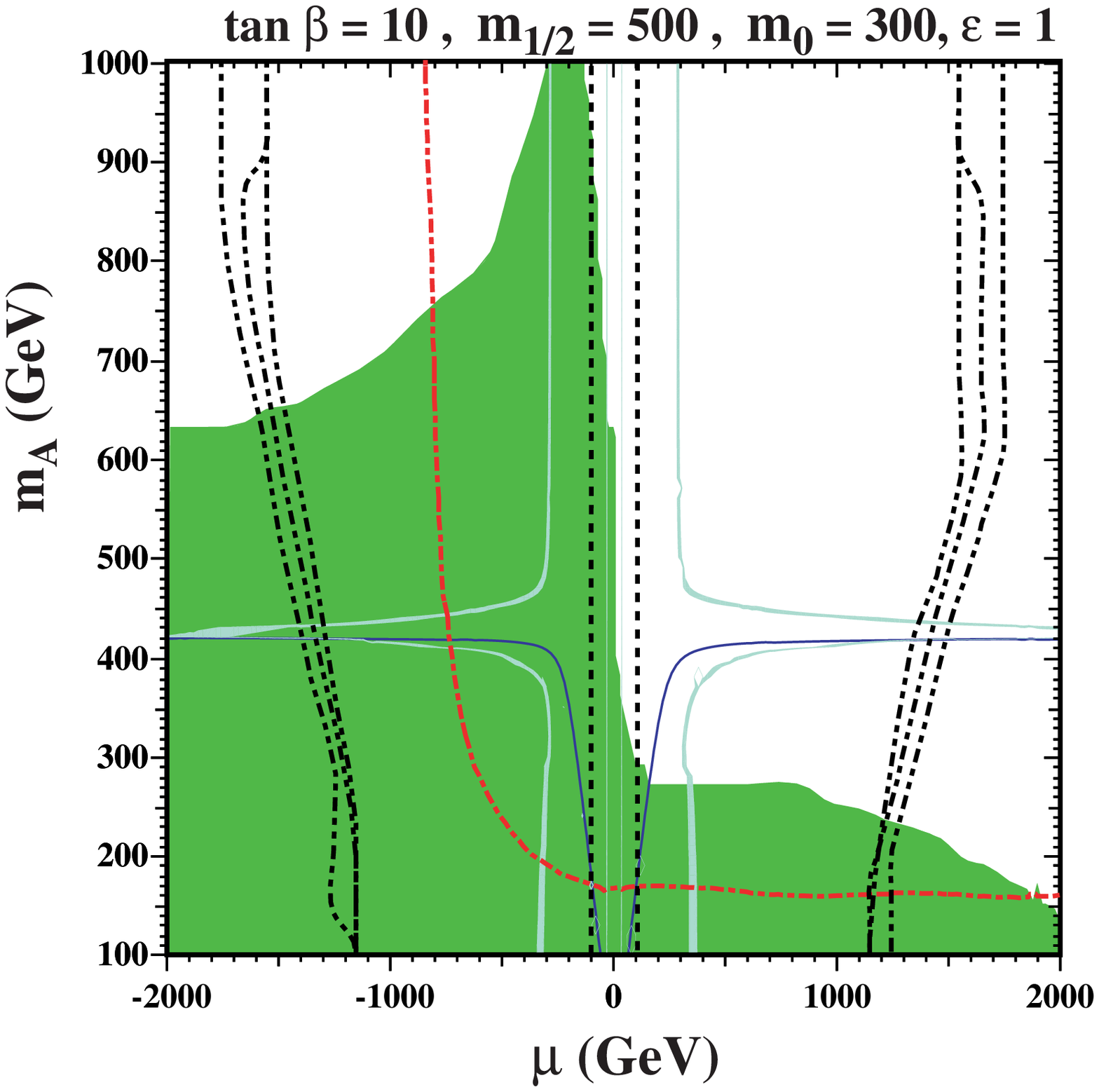} 
& \includegraphics[width=2.7in,height=2.7in]{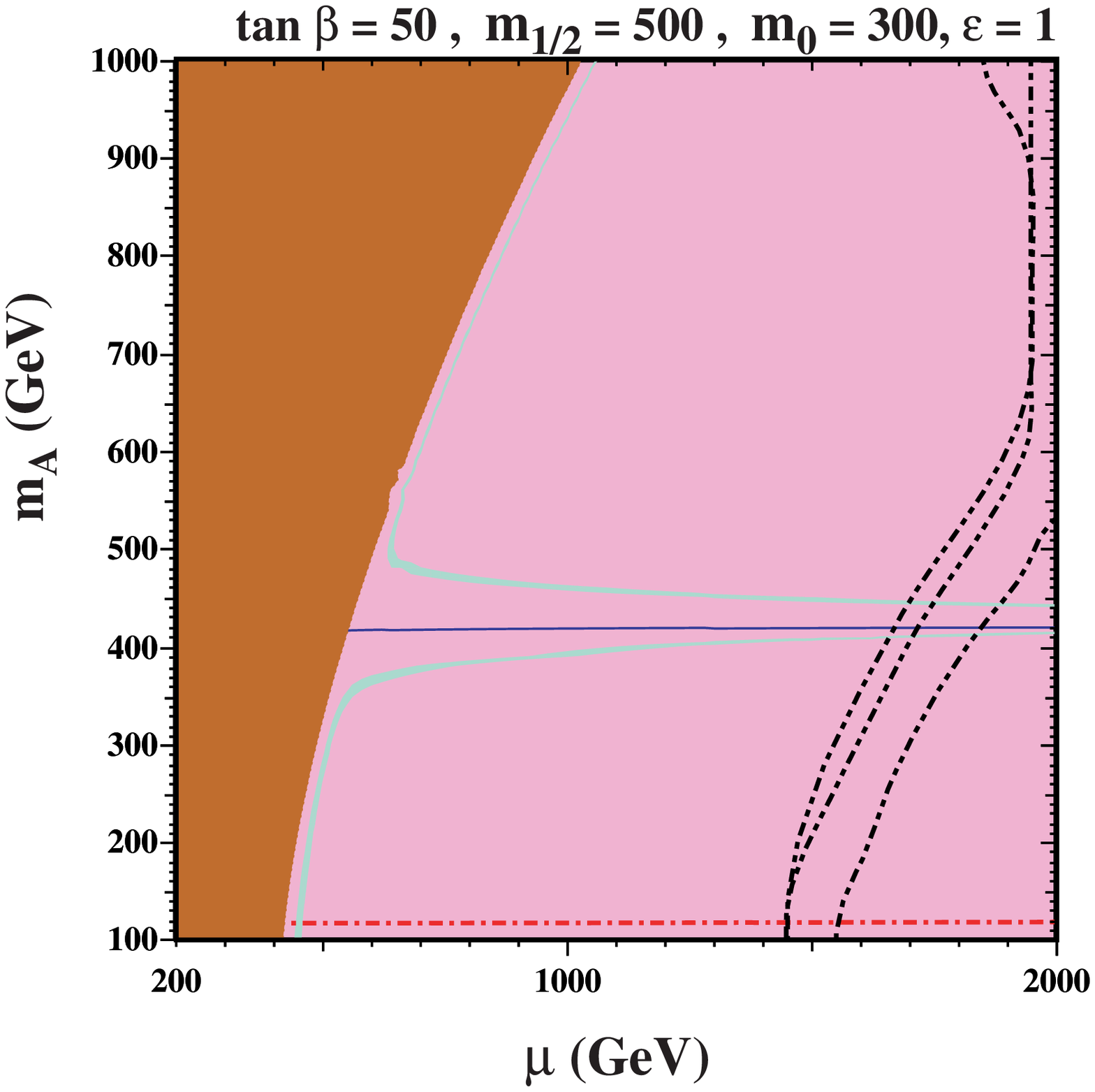} \\
\includegraphics[width=2.7in,height=2.7in]{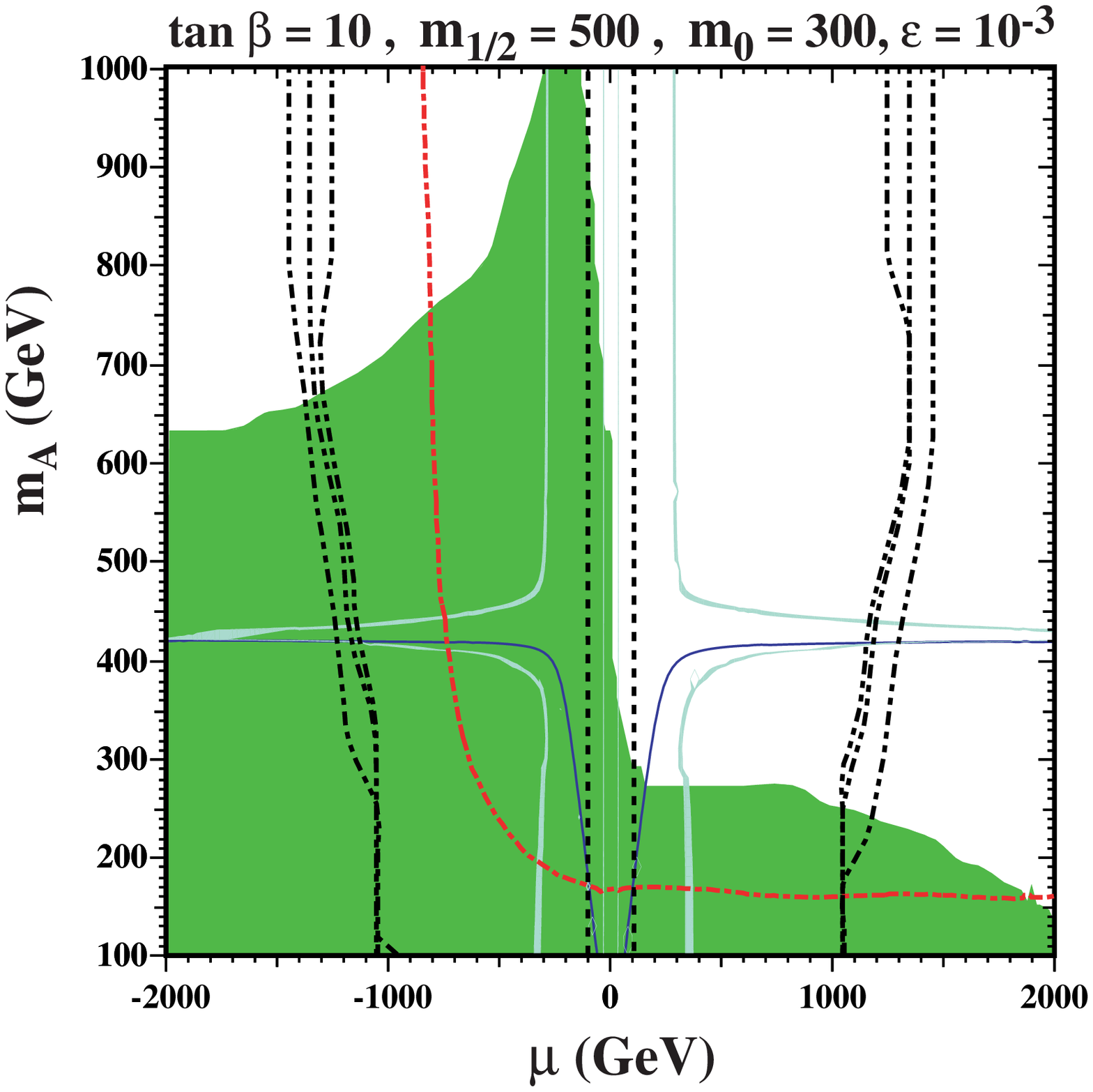} 
& \includegraphics[width=2.7in,height=2.7in]{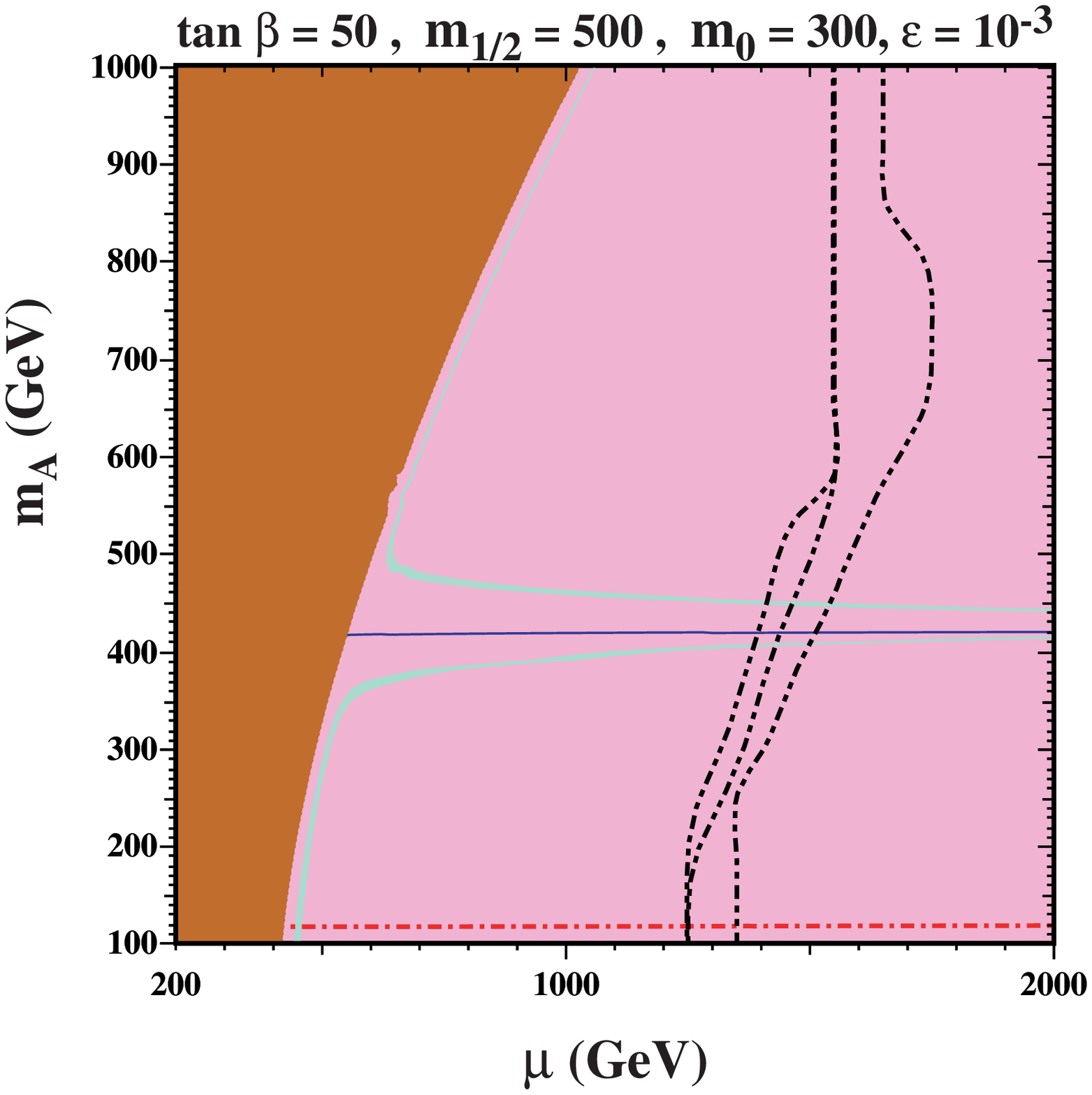} \\
\includegraphics[width=2.7in,height=2.7in]{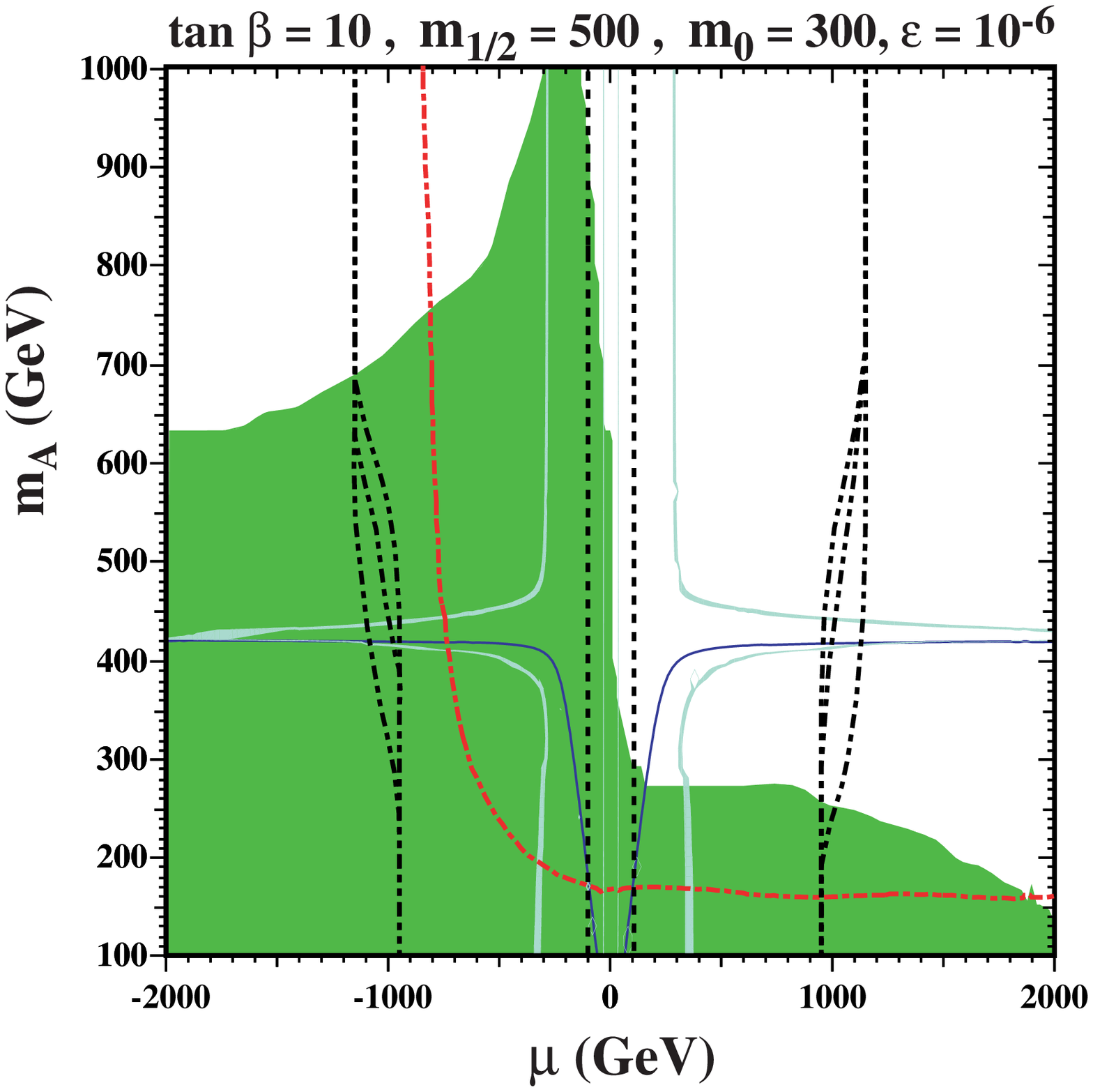} 
& \includegraphics[width=2.7in,height=2.7in]{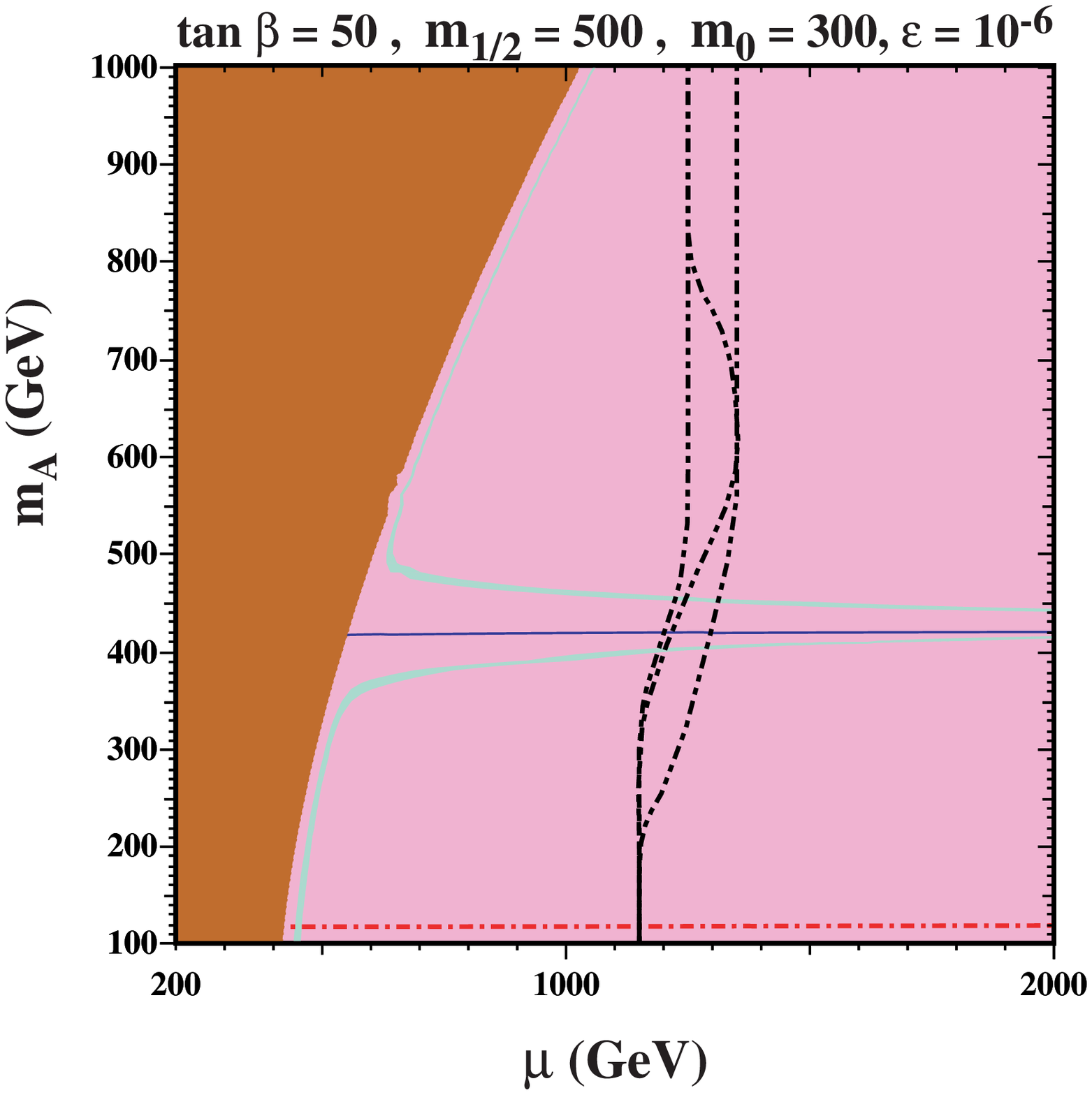} \\
\end{tabular}
\caption{\it Portions of the NUHM $(\mu, m_A)$ planes for for  $m_{1/2} = 500$~GeV, 
$m_0 = 300$~GeV,
$A_0 =0$ and (a, c, e) $\tan \beta = 10$, (b, d, f) $\tan \beta = 50$. The notations are 
described in the text. The regions outside the inner dash-dotted curves have problematic 
high-scale tachyons for higher values of $\mfund$.
\label{vef2}
}
\end{center}
\end{figure}

In Fig.~\ref{vef2}, we show more examples of $(\mu, m_A)$ planes,
this time with $m_{1/2} = 500$ GeV, $m_0 = 300$ GeV, and $A_0 = 0$.
We see that there are regions at larger $m_A$, particularly for $\mu > 0$,
where the LEP Higgs constraint is respected. 
As in Fig.~\ref{vef1}, the problematic tachyonic regions extend
to smaller $|\mu|$ as $m_A$ decreases, and as $\epsilon$ decreases from
1 to $10^{-6}$. The potential tachyonic problem is restricted essentially to
$|\mu| \gappeq 1000$~GeV.

Turning to panels (b), (d) and (f) in Fig.~\ref{vef2}, for $\tan \beta = 50$,
we seethat the stau LSP region extends to between $\mu \sim 400$~GeV (for
small $m_A$) and $\mu \sim 1000$~GeV (for large $m_A$).
The problematic tachyon region is now only at larger $\mu$, varying between
a range $> 2000$~GeV for $\epsilon = 1$ and large $m_A$ to $\sim 1150$~GeV
for $\epsilon = 10^{-6}$ and small $m_A$. In all cases, the portion of the WMAP
strip where the relic neutralino density is controlled by stau coannihilation
is in the safe region, as well as a portion of the rapid-annihilation funnel
where $m_\chi \sim m_A/2$. Whether the other regions are acceptable would
depend on the cosmological scenario. Note that the entire regions shown in panels
(b), (d) and (f) are favoured by $g_\mu - 2$.

Comparing Figs.~\ref{vef1} and \ref{vef2}, we see (unsurprisingly)
that the problematic tachyonic NUHM regions grow as $m_0$ decreases.
As we noted above, it is also possible to consider $m_0^2 < 0$ in the NUHM.
For example, For $m_{1/2} = 300$ GeV and $m_0^2/\sqrt{|m_0^2|}$  = -100 GeV,
with $\mfund = M_{GUT}$ and $\epsilon = 1$, CCB vacua appear at high scale when
$|\mu| > 1000$  GeV for $m_A \gappeq 800$ GeV and $|\mu| > 1500$ GeV for lower $m_A$.
When $\epsilon = 10^{-3}$ for the same case, the entire plane is problematic.
Similarly, when $\mfund = M_P$ for any value of $\epsilon$, the entire plane is 
problematic.

\section{Summary}

We have discussed in this paper the constraints that are imposed on
supersymmetric models by the presence of the Universe in our familiar EW vacuum.
We have argued that models with tachyonic spin-zero
fields at some high input scale are not necessarily excluded.
The renormalization-group evolution of tachyonic masses to low scales
may change their signs, in which case the standard EW vacuum would be a
local minimum of the effective potential. However,
in addition to this vacuum, there may be a much deeper high-scale vacuum located along 
some F- and D-flat direction in the effective potential, with field values fixed by
some higher-order non-renormalizable interaction. We discuss the circumstances
under which the existence and location of such a high-scale vacuum can be
calculated reliably.

Such high-scale vacua usually
break both color and charge conservation. In general, the lifetime for decay of the EW
vacuum to this unacceptable lower minimum of the effective potential is much longer than 
the age of the Universe, so future decay into such a vacuum is not of immediate concern.
A more relevant question is whether the Universe would have fallen into such
a vacuum during its past history. This depends whether the effective scalar
masses-squared acquired large positive or negative contributions $\propto H^2$ during inflation.
If these contributions were ${\cal O}(1)$ and positive, only the EW vacuum
would be populated. On the other hand if these contributions were negative
(and possibly if they were positive but small), the high-scale vacuum would be populated.

We have then explored the conditions under which the CMSSM or the NUHM
(with its two extra parameters) has a calculable high-scale vacuum. If these
conditions are not satisfied, there is no reason to be tachyophobic. Even if these
conditions are satisfied, and there is a calculable high-scale vacuum, whether it
is catastrophic or not depends on early cosmology, and there is still no need
to be tachyophobic.

Unreasoning tachyophobia is never justified: one must examine rationally
whether any specific tachyonic spin-zero field is dangerous, depending on the evolution
of the Universe within one's favoured cosmological scenario.

\appendix

\section{Generalities}
\label{gen}
Here we give a concise review of the well-known
nature of the flat directions that will be
considered.  This Appendix is included to serve as a reminder and 
to set our notation.

We begin our discussion by ignoring the superpotential
and soft terms, which are included later.
We assume complex scalar fields $\phi_i$ with a
canonical K\"ahler potential, with vevs denoted by $v_i$.
We take $v_i$ to correspond to a supersymmetric
vacuum, $D^a = 0$, in the standard notation.
Viewed as a vector space, the $v_i$ are the 
null vectors of the (hermitian) mass-squared matrix:
\beq
\sum_k \bar v_k M_{k \bar \ell}^2 = \sum_{k,a} g_a^2\frac{\p D^a}{\p v_k}
\frac{\p D^a}{\p \bar v_{\bar \ell}} \bar v_k = 
\sum_a g_a^2 D^a \frac{\p D^a}{\p \bar v_{\bar \ell}} = 0.
\eeq
Correspondingly, there exists a projection
operator into this null space:
\beq
P_{ij} = \frac{v_i \bar v_j}{\sum_k |v_k|^2}.
\eeq
which may be used to construct the modulus field $\eta$, which
is the scalar field tangential to the null space
defined by $P$:
\beq
\eta = \sum_j \frac{\bar v_j}{\sqrt{\sum_k |v_k|^2}} \phi_j.
\label{edf}
\eeq
In addition, there are the other modes $\chi_\alpha$, that
are orthogonal to $\eta$ in the space of the $\phi_i$, i.e.:
\beq
\chi_\alpha = \sum_i \bar \theta_i^\alpha \phi_i : \quad
\sum_i \theta_i^\alpha \bar v_i =0, \quad
\sum_i \bar \theta_i^\alpha \theta_i^\beta = \delta^{\alpha \beta}.
\label{omd}
\eeq
We now consider the effect of the soft masses on the
modulus field $\eta$.
When the change of variables  \myref{edf} is made, we obtain:
\beq
V_{\text{soft}} \ni \sum_i \tmsq_i |\phi_i|^2
\ni \frac{\sum_i |v_i|^2 \tmsq_i}{\sum_k |v_k|^2} |\eta|^2
\equiv \tmsq |\eta|^2 .
\label{mtd}
\eeq
In the case of a flat direction characterized by a
single monomial
\beq
\Phi = \prod_i \phi_i^{n_i},
\label{mfd}
\eeq
the D-flatness constraint 
\beq
\bar \phi_i = \frac{\p \Phi}{\p \phi_i} \cdot \text{const}
\quad \forall \quad i
\eeq
yields:
\beq
|v_i|^2 = n_i \cdot \text{const} \quad \forall \quad i .
\eeq
It follows that in this case the modulus soft mass-squared 
appearing in \myref{mtd} is:
\beq
\tmsq = \frac{\sum_i n_i \tmsq_i}{\sum_k n_k} ,
\label{hyt}
\eeq
where we see that the weights are nothing but the powers in
the monomial.  Because we specialize to the case
of monomial flat directions \myref{mfd} in our study,
we use \myref{hyt} to determine the soft masses of moduli.  
The cases that we are interested in
are those with 
\beq
\tmsq(Q)<0 ,
\label{tach}
\eeq
where $Q$ is the running scale.  In such a case, the modulus runs away from the origin
along the flat direction.

Next we introduce the non-renormalizable superpotential
term that lifts the flat direction at large field values
and stabilizes the modulus against the runaway behavior:
\beq
W_{NR} = \frac{ 1 }{\mfund^{n-3}} \eta^{n-1} \sum_\alpha s_\alpha \chi_\alpha
+ \frac{t}{n \mfund^{n-3}} \eta^n.
\label{nrsp}
\eeq
Here $s_\alpha,t$ are coupling strengths.
The fields $\chi_\alpha$ represent the non-modulus modes \myref{omd}.
Below, we make the simplifying assumption that only a single flat
direction is `turned on', so that $\vev{\chi_\alpha}=0$.

In Appendix \ref{lift} we provide examples
that yield the two types of terms appearing
in \myref{nrsp}.  We remark that manifest gauge invariance
is typically lost when we use the basis $\eta,\chi_\alpha$,
as reflected in the appearance of the $\eta^n$ coupling.  
As illustrated by the examples provided in the
Appendix \ref{lift}, the power law dependences on $\tmsq$ and $\mfund$ 
exhibited in \myref{pwr} are quite generic.

The non-renormalizable superpotential \myref{nrsp}
may well be the result of the exchanges
of states with mass scale $\mfund$ that have
been integrated out. The scale $\mfund$ could be as low as the
GUT scale, if appropriate GUT representations
are coupled to the MSSM content.  Alternatively, $\mfund$
could be as high as the Planck scale, if it is
due to the exchange of quantum-gravitational excitations.
Another possibility is that $\mfund \approx
M_s \approx 4$ to $5 \times 10^{17}$ GeV, the
perturbative heterotic string scale.

Whatever $\mfund$ happens to be,
the scalar potential for the modulus $\eta$ is given
by \myref{potnr}.  One finds that
$|u|^2 \equiv |t|^2 + \sum_\alpha |s_\alpha|^2$ combines the coupling
constants appearing in \myref{nrsp}.
Since $\tmsq<0$, the minimum is obtained at:
\beq
|\eta| = \[ \frac{ -\tmsq \mfund^{2(n-3)}}{|u|^2 (n-1)} \]^{\frac{1}{2(n-2)}}.
\label{exan}
\eeq
In the numerical analysis of flat directions
that we perform, we specify the vev $v$ of the modulus
$\eta$ according to the power law that has just been
obtained, namely \myref{pwr}.
As we have commented above, this estimate is
consistent with a detailed analysis of the non-renormalizable
interactions that are allowed in the MSSM.
For our purposes, such an order of magnitude estimate is sufficient .

\section{Examples of flat direction lifts}
\label{lift}
\subsection{Lift of the $H_u H_d$ flat direction}
The leading non-renormalizable superpotential term
that achieves this is
\beq
W_{NR} = \frac{t}{M_*} (H_u H_d)^2,
\eeq
where $t$ is a coupling constant.
We choose the variant of the flat direction for which
the neutral components $H_u^0$ and $H_d^0$
get vevs that are equal.  In that case the modulus $\eta$
and orthogonal modes $\chi_\alpha$ are:
\beq
\eta = \frac{1}{\sqtw} (H_u^0+H_d^0), \quad
\chi_1 = \frac{1}{\sqtw} (H_u^0-H_d^0), \quad
\chi_2 = H_u^+, \quad \chi_3=H_d^- .
\eeq
Using these redefinitions, we find that:
\beq
\tilde W_{NR} = \frac{t}{4M_*} \eta^4 + \ord{\chi^2} ,
\eeq
where symmetries forbid $\chi_\alpha \eta^3$ terms.  One such symmetry is
$H_u^0 \leftrightarrow H_d^0$,
which imposes symmetry under $\chi_1 \to -\chi_1$ with $\eta,\chi_2,\chi_3$
invariant, which forbids a $\chi_1 \eta^3$ term.  The other symmetry is
$H_u^0 \to -H_u^0, H_d^0 \to -H_d^0$, which imposes symmetry under
$\eta \to -\eta$, $\chi_1 \to -\chi_1$ with $\chi_2,\chi_3$
invariant.  This symmetry forbids $\chi_2 \eta^3$ and $\chi_3 \eta^3$.
Thus we see that the coefficients $s_\alpha$ in \myref{nrsp}
vanish for the present case.  The formula \myref{exan}
applies for the vev, with $u \to t$ and $n=4$.  Note that
$\tmsq = |\mu|^2 + (\tmsq_{H_u} + \tmsq_{H_d})/2 < 0$
must hold for this vev  to run away, where $\mu$
is as usual the (tree-level) MSSM higgsino mixing term.

\subsection{Lift of the $L_1 L_2 \tau^c$ flat direction}
We choose the direction where the vevs of
\beq
\nu_e = \mu = \tau^c = v, \quad \nu_\mu = e = 0.
\eeq
The leading non-renormalizable superpotential
term that lifts this flat direction is
\beq
W_{NR} = -\frac{9 s}{M_*^2} (H_u L_2) (L_1 L_2) \tau^c,
\eeq
where the parentheses indicate the SU(2) invariants
and the coefficient has been selected to provide a
simple final answer, as becomes evident shortly.
As usual, $s$ is a dimensionless coupling constant.
The modulus $\eta$ and orthogonal modes $\chi_\alpha$
can be parameterized as:
\beq
\eta=\frac{1}{\sqrt{3}} ( \nu_e + \mu + \tau^c ), \quad
\chi_1 = \frac{1}{\sqtw} ( \nu_e - \mu), \quad 
\chi_2 = \frac{1}{\sqrt{6}} ( \nu_e + \mu - 2 \tau^c), \quad
\chi_3 = H_u^0,
\eeq
with $\chi_{4,5,6}$ corresponding to the other
superfields $\nu_\mu, e, H_u^+$.
One finds:
\beq
W_{NR} = \frac{s}{M_*^2} \chi_3 \eta^4 + \ord{\chi^2}.
\eeq
Thus the stabilization term is of the form
of the first term in \myref{nrsp}.  The $\eta^5$
term, with coefficient $t$, is forbidden by matter
parity.  Symmetry arguments also forbid $\chi_\alpha \eta^4$
with $\alpha=1,2,4,5,6$.
The formula \myref{exan}
applies for the vev, with $u \to s$ and $n=5$.

\section{Method of scan}

In this appendix we summarize the workflow used for our analysis of
flat directions.

\ben
\item
Starting from unification-scale boundary conditions $m_0, M_{1/2},
A_0$,  and electroweak scale $\tan\beta$,
we evolve the RGEs to the EW scale, iterating $\mu$ and $B\mu$ until
valid electroweak symmetry breaking (EWSB) is obtained.
We exclude automatically any model with tachyonic squarks or sleptons at 
the low scale.
\item
Next, we loop through all flat directions enumerated
in~\cite{GKM} (limiting ourselves to monomials \myref{mfd}) and
perform an analysis over the range of running scales $Q \geq 10$ TeV (see
the discussion in Section \ref{tnu}):
 \ben
 \item
 We check whether $\tmsq<0$ for the weighted sum \myref{hyt}.  The powers
 $n_i$ and masses $\tmsq_i$ that appear in this sum depend on the
 flat direction that is chosen; because the mass parameters depend
 on $Q$, so does the weighted sum $\tmsq$;
 \item
 For values of $Q$ such that $\tmsq<0$, we determine the corresponding
 vev $v$ using~\myref{pwr};
 \item
 We then compare $v$ to $\e Q$ for various values of $\e$:
   \bit
   \item
   If $v\leq \e Q$ for all $Q$, our analysis does not show the existence of a CCB minimum;
   \item
   Otherwise, the CCB vacuum does exist  since the RG-improved analysis
   tells us that there is a potential 
   minimum at large $v$ and the one-loop corrections
   are small over some range of $Q$ for which the vev is nonzero.
   \eit
 \een
\een



\begin{thebibliography}{9}

\bibitem{cmssmwmap}
  J.~R.~Ellis, K.~A.~Olive, Y.~Santoso and V.~C.~Spanos,
  Phys.\ Lett.\  B {\bf 565}, 176 (2003)
  [arXiv:hep-ph/0303043];
  H.~Baer and C.~Balazs,
  JCAP {\bf 0305}, 006 (2003)
  [arXiv:hep-ph/0303114];
  A.~B.~Lahanas and D.~V.~Nanopoulos,
  Phys.\ Lett.\  B {\bf 568}, 55 (2003)
  [arXiv:hep-ph/0303130];
  U.~Chattopadhyay, A.~Corsetti and P.~Nath,
  Phys.\ Rev.\  D {\bf 68}, 035005 (2003)
  [arXiv:hep-ph/0303201];
  C.~Munoz,
  Int.\ J.\ Mod.\ Phys.\  A {\bf 19}, 3093 (2004)
  [arXiv:hep-ph/0309346].

 \bibitem{gdm}
  J.~R.~Ellis, K.~A.~Olive, Y.~Santoso and V.~C.~Spanos,
  Phys.\ Lett.\ B {\bf 588} (2004) 7
  [arXiv:hep-ph/0312262];
J.~L.~Feng, A.~Rajaraman and F.~Takayama,
  Phys.\ Rev.\ Lett.\  {\bf 91} (2003) 011302
  [arXiv:hep-ph/0302215];
  J.~L.~Feng, S.~Su and F.~Takayama,
  Phys.\ Rev.\ D {\bf 70} (2004) 075019
  [arXiv:hep-ph/0404231].
  
  \bibitem{vcmssm}
J.~R.~Ellis, K.~A.~Olive, Y.~Santoso and V.~C.~Spanos,
  Phys.\ Lett.\ B {\bf 573} (2003) 162
  [arXiv:hep-ph/0305212];
  J.~R.~Ellis, K.~A.~Olive, Y.~Santoso and V.~C.~Spanos,
  Phys.\ Rev.\ D {\bf 70} (2004) 055005
  [arXiv:hep-ph/0405110].

\bibitem{m2neg}
 J.~L.~Feng, A.~Rajaraman and B.~T.~Smith,
  Phys.\ Rev.\  D {\bf 74}, 015013 (2006)
  [arXiv:hep-ph/0512172];
 A.~Rajaraman and B.~T.~Smith,
  Phys.\ Rev.\  D {\bf 75}, 115015 (2007)
  [arXiv:hep-ph/0612235].

\bibitem{Lebedev:2005ge}
  O.~Lebedev, H.~P.~Nilles and M.~Ratz,
  arXiv:hep-ph/0511320.


\bibitem{Choi:2005hd}
  K.~Choi, K.~S.~Jeong, T.~Kobayashi and K.~i.~Okumura,
  Phys.\ Lett.\  B {\bf 633}, 355 (2006)
  [arXiv:hep-ph/0508029];
R.~Kitano and Y.~Nomura,
  Phys.\ Lett.\  B {\bf 631}, 58 (2005)
  [arXiv:hep-ph/0509039].


  
\bibitem{ccb}
G.~Gamberini, G.~Ridolfi and F.~Zwirner,
  Nucl.\ Phys.\  B {\bf 331}, 331 (1990);
J.~A.~Casas, A.~Lleyda and C.~Munoz,
Nucl.\ Phys.\ B {\bf 471} (1996) 3
[arXiv:hep-ph/9507294];
and
Phys.\ Lett.\ B {\bf 389} (1996) 305
[arXiv:hep-ph/9606212];
A.~Strumia,
Nucl.\ Phys.\ B {\bf 482} (1996) 24
[arXiv:hep-ph/9604417];
H.~Baer, M.~Brhlik and D.~Castano,
Phys.\ Rev.\ D {\bf 54} (1996) 6944
[arXiv:hep-ph/9607465];
S.~A.~Abel and C.~A.~Savoy,
Nucl.\ Phys.\ B {\bf 532} (1998) 3
[arXiv:hep-ph/9803218];
S.~Abel and T.~Falk,
Phys.\ Lett.\ B {\bf 444} (1998) 427
[arXiv:hep-ph/9810297];
D.~G.~Cerdeno, E.~Gabrielli, M.~E.~Gomez and C.~Munoz,
JHEP {\bf 0306} (2003) 030
[arXiv:hep-ph/0304115].

  \bibitem{FORS}
T.~Falk, K.~A.~Olive, L.~Roszkowski and M.~Srednicki,
Phys.\ Lett.\ B {\bf 367} (1996) 183
[arXiv:hep-ph/9510308].

\bibitem{RR}
A.~Riotto and E.~Roulet,
Phys.\ Lett.\ B {\bf 377} (1996) 60
[arXiv:hep-ph/9512401];
A.~Kusenko, P.~Langacker and G.~Segre,
Phys.\ Rev.\ D {\bf 54} (1996) 5824
[arXiv:hep-ph/9602414].

\bibitem{AD}
I.~Affleck and M.~Dine,
Nucl.\ Phys.\  B {\bf 249}, 361 (1985).


\bibitem{FORSS}
T.~Falk, K.~A.~Olive, L.~Roszkowski, A.~Singh and M.~Srednicki,
Phys.\ Lett.\ B {\bf 396} (1997) 50
[arXiv:hep-ph/9611325].

  \bibitem{nonu} 
  M.~Olechowski and S.~Pokorski,
  Phys.\ Lett.\ B {\bf 344}, 201 (1995)
  [arXiv:hep-ph/9407404];
V.~Berezinsky, A.~Bottino, J.~Ellis, N.~Fornengo, 
               G.~Mignola and S.~Scopel,
               {\em Astropart.\ Phys.}  {\bf 5} (1996) 1, 
               hep-ph/9508249;
               M.~Drees, M.~Nojiri, D.~Roy and Y.~Yamada,
               {\em Phys.\ Rev.} {\bf D 56} (1997) 276, 
               [Erratum-ibid.\ {\bf D 64} (1997) 039901], 
               hep-ph/9701219;
               M.~Drees, Y.~Kim, M.~Nojiri, D.~Toya, K.~Hasuko and 
               T.~Kobayashi,
               {\em Phys.\ Rev.} {\bf D 63} (2001) 035008, 
               hep-ph/0007202;
               P.~Nath and R.~Arnowitt,
               {\em Phys.\ Rev.} {\bf D 56} (1997) 2820, 
               hep-ph/9701301;
               J.~R.~Ellis, T.~Falk, G.~Ganis, K.~A.~Olive and M.~Schmitt,
  Phys.\ Rev.\ D {\bf 58} (1998) 095002
  [arXiv:hep-ph/9801445];
J.~R.~Ellis, T.~Falk, G.~Ganis and K.~A.~Olive,
  Phys.\ Rev.\ D {\bf 62} (2000) 075010
  [arXiv:hep-ph/0004169];
               A.~Bottino, F.~Donato, N.~Fornengo and S.~Scopel,
               {\em Phys.\ Rev.} {\bf D 63} (2001) 125003, 
               hep-ph/0010203;
               S.~Profumo,
               {\em Phys.\ Rev.} {\bf D 68} (2003) 015006, 
               hep-ph/0304071;
               D.~Cerdeno and C.~Munoz,
               {\em JHEP} {\bf 0410} (2004) 015, 
               hep-ph/0405057;
               H.~Baer, A.~Mustafayev, S.~Profumo, A.~Belyaev and X.~Tata,
               {\em JHEP} {\bf 0507} (2005) 065, 
               hep-ph/0504001.


\bibitem{nuhm}
J.~Ellis, K.~Olive and Y.~Santoso,
Phys. Lett.  B~{\bf 539}, 107 (2002)
[arXiv:hep-ph/0204192];
J.~R.~Ellis, T.~Falk, K.~A.~Olive and Y.~Santoso,
Nucl.\ Phys.\ B {\bf 652}, 259 (2003)
[arXiv:hep-ph/0210205].


\bibitem{GKM}
T.~Gherghetta, C.~F.~Kolda and S.~P.~Martin,
Nucl.\ Phys.\ B {\bf 468} (1996) 37
[arXiv:hep-ph/9510370].


\bibitem{lin2}A.D. Linde, Phys. Lett. {\bf 116B} (1982) 335;
A.A. Starobinsky, Phys. Lett. {\bf 117B} (1982) 175;
A. Vilenkin, Nucl. Phys. {\bf B226} (1983) 527;
K. Enqvist, K.W. Ng, and K.A. Olive, Nucl. Phys. {\bf B303} (1988) 713.

\bibitem{drt} M.Dine, L. Randall, and S. Thomas, Phys. Rev. Lett. {\bf
    75} (1995) 398; Nucl. Phys. {\bf B458} (1996) 291.
    
    \bibitem{gmo} M.~K. Gaillard, H. Murayama, and K. Olive, Phys. Lett.
{\bf B355} (1995) 71;
 B.~A.~Campbell, M.~K.~Gaillard, H.~Murayama and K.~A.~Olive,
  Nucl.\ Phys.\ B {\bf 538}, 351 (1999)
  [arXiv:hep-ph/9805300].



\bibitem{BG} P.~Binetruy and M.K.~Gaillard, 
Phys. Lett. {\bf B195} (1987) 382.

 \bibitem{ns} E. Cremmer, S. Ferrara, C. Kounnas and D.V. Nanopoulos, Phys. Lett.
    {\bf 133B} (1983) 61.

\bibitem{cefos}
R.~H.~Cyburt, J.~R.~Ellis, B.~D.~Fields, K.~A.~Olive and V.~C.~Spanos,
  JCAP {\bf 0611}, 014 (2006)
  [arXiv:astro-ph/0608562].



\bibitem{FeynHiggs}
S.~Heinemeyer, W.~Hollik and G.~Weiglein,
{\it Comput.\ Phys.\ Commun.\ } {\bf 124} (2000) 76 
[arXiv:hep-ph/9812320];
S.~Heinemeyer, W.~Hollik and G.~Weiglein,
{\it Eur.\ Phys.\ J.\ C} {\bf 9} (1999) 343 
[arXiv:hep-ph/9812472].

\bibitem{wmap}
D.~N.~Spergel {\it et al.}  [WMAP Collaboration],
  Astrophys.\ J.\ Suppl.\  {\bf 170} (2007) 377
  [arXiv:astro-ph/0603449].


\end{thebibliography}
\end{document}